\documentclass[aps,pre,amsmath,amsfonts,amssymb,floatfix,superscriptaddress,nofootinbib]{revtex4}
\usepackage[latin1]{inputenc}
\usepackage{subfigure}
\usepackage{float}
\usepackage{amsmath}
\usepackage[english]{babel}
\usepackage{amsfonts}
\usepackage{amssymb}
\usepackage{fancyhdr}
\usepackage{esdiff}
\usepackage{comment}
\usepackage{textcomp}
\usepackage{cancel}
\usepackage{graphicx}
\usepackage[font=small]{caption}
\usepackage{epsfig}
\usepackage{epstopdf}
\usepackage[T1]{fontenc}
\usepackage{lmodern}

\begin{document}
	\title{Spectral control for ecological stability}

	\author{Giulia Cencetti} 
	\affiliation{Dipartimento di Ingegneria dell'Informazione, Universit\`{a} di Firenze,
		Via S. Marta 3, 50139 Florence, Italy}
	\affiliation{INFN Sezione di Firenze, via G. Sansone 1, 50019 Sesto Fiorentino, Italia}
	\author{Franco Bagnoli}  \affiliation{Universit\`{a} degli Studi di Firenze, Dipartimento di Fisica e Astronomia and CSDC, via G. Sansone 1, 50019 Sesto Fiorentino, Italia}
	\affiliation{INFN Sezione di Firenze, via G. Sansone 1, 50019 Sesto Fiorentino, Italia}
	\author{Giorgio Battistelli} 
	\affiliation{Dipartimento di Ingegneria dell'Informazione, Universit\`{a} di Firenze,
		Via S. Marta 3, 50139 Florence, Italy}
	\author{Luigi Chisci} 
	\affiliation{Dipartimento di Ingegneria dell'Informazione, Universit\`{a} di Firenze,
		Via S. Marta 3, 50139 Florence, Italy}
	\author{Duccio Fanelli} \affiliation{Universit\`{a} degli Studi di Firenze, Dipartimento di Fisica e Astronomia and CSDC, via G. Sansone 1, 50019 Sesto Fiorentino, Italia}
	\affiliation{INFN Sezione di Firenze, via G. Sansone 1, 50019 Sesto Fiorentino, Italia}

	\begin{abstract}
			A system made up of $N$ interacting species is considered.  Self-reaction terms are assumed of the logistic type. Pairwise interactions take place among species according to different modalities, thus yielding a complex asymmetric disordered graph. A mathematical procedure is introduced and tested to stabilise the ecosystem via an {\it ad hoc} rewiring of the underlying couplings. The method implements minimal modifications to the spectrum of the Jacobian matrix which sets the stability of the fixed point and traces these changes back to species-species interactions. Resilience of the equilibrium state appear to be favoured by predator-prey interactions.

	\end{abstract}

	\maketitle

\section{Introduction}


A vast plethora of physical phenomena occurring in nature are due to collective dynamics, which spontaneously emerge at the macroscopic level in systems made up of microscopically interacting constituents. When the entities that compose the whole set are subject to specific self-reactions and, at the same time, diffuse across the embedding spatial medium, the system goes under the name of reaction-diffusion  \cite{Turing52, Murray02}. Examples are invasion models in ecology \cite{HolmesLewisBanksVeit94}, epidemic spreading \cite{MurrayStanleyBrwon86}, and also the celebrated Turing patterns that arise, for instance, from the dynamical interplay between reaction and diffusion in a chemical system \cite{Turing52}.
Usually, reaction-diffusion models are defined on a regular lattice, either continuous or discrete \cite{Murray02, Smoller94, Grindrod96}. In many cases of interest, it is however necessary to consider the units moving on a complex network \cite{OthmerScriven71, OthmerScriven74, NakaoMikhailov10, AsllaniChallengerPavoneSacconiFanelli14, AsllaniDiPattiFanelli12, AngstmannDonnellyHenry13}, where each node represents a physical location; individual elements are constantly displaced from one node to another following diffusion rules. Spatial ecological systems can be schematized by resorting to a network that exemplifies existing routes linking different habitats, infection spreading customarily requires accounting for a transportation network, and also chemical reactors and biological cells can be considered as coupled through complex networks.\\
A reaction-diffusion model defined on a complex network can display a stationary stable equilibrium. Stability is of paramount importance as it relates to resilience, the ability of the system to counteract external perturbations that would tend to get away from the existing equilibrium. It is therefore crucial to devise possible strategies aimed at interfering with the system of interest so as to enforce the desired stability. This can be achieved by acting on the local dynamics or, more interestingly for what it follows, by reshaping the underlying network of spatial connections. This latter possibility has been addressed in \cite{Cencetti_etal17}, via a minimally invasive procedure targeted to modifying the spectrum of the Laplacian operator which governs the diffusive, hence linear, couplings. This, in turn, implies rewiring the links, and consequently recalibrating the weights, of the underlying graph of connections.\\ 
Dynamics on networks is however central to other realms of investigation, also when the system being examined is not made spatially extended. In this case, long-ranged (non local) interactions, as encoded in the structure of the assigned network, might follow {\it non-diffusive} rules.  In ecosystems inter-species interactions are assumed to be 
mediated by pairwise, hence quadratic, exchange, to some level of approximation. Each population (species) is characterised by a self-reaction dynamics, typically described via a suitable nonlinear function of the concentration amounts, and different populations can be abstractly assigned to given nodes of a virtual graph. The network then represents the interactions between different species and the sign of the weighted entries of the associated adjacency matrix define the nature of coupling (competitive, cooperative, predator-prey, etc...).\\
The stability of the ecosystem, or its resilience, is essential in many respects. In the past decades many efforts have been made to understand the principles that rule the stability of a complex ecosystem. These concepts have been originally addressed by Robert May \cite{May72} in a seminal paper that paved the way to a completely new field of investigation, still very productive and fertile. These studies resulted in multiple attempts of providing methods to make an ecological community structurally stable \cite{GaoBarzelBarabasi16,SuweisSiminiBanavarMaritan13,SuweisGrilliMaritan14,SuweisGrilliBanavarAllesinaMaritan15,Grilli_etal17,ThebaultFontaine10,Johnson_etal14,Vazquez_etal07}. May \cite{May72} analysed a system described by $N$ variables ($N$ interacting species) obeying a set of differential equations. The stability analysis is performed by linearising the equations in the neighbourhood of an equilibrium point, whose stability depends on the spectrum of the interaction matrix. May's analysis focuses on this latter, eventually bearing to the challenging statement that, in short, the more complex the more unstable is the system. Recent work by Allesina et al. \cite{AllesinaTang12} provides an implementation of May's ecosystem accounting for well defined (non random) interactions. These are competitive, mutualistic and predator-prey and according to Allesina et al. play different roles in the stability of the ecosystem. Remarkably, the presence of predator-prey relations has a stabilizing effect. In Coyte et al. \cite{CoyteSchluterFoster15} stability of a microbioma ecosystem is obtained by allowing for sufficiently weak couplings. \\
Starting from these premises, in this paper we provide an alternative approach to ecological stability by developing a self-consistent mathematical strategy which implements a spectral control algorithm. The method builds on the technique developed in \cite{Cencetti_etal17} and extends its domain of applicability, beyond diffusion mediated (linear) processes to the interesting setting where pairwise, hence non-linear, non-local interactions are considered. In doing so, we will contribute to identifying the key topological features that should be possessed by a stable (resilient) ecological network. To anticipate our findings we will show that predator-prey interactions exert a beneficial role in terms of stability in qualitative agreement with the results reported in  \cite{AllesinaTang12}. Further weak interactions tend to favour the overall stabilization, as observed in \cite{CoyteSchluterFoster15}.\\
The paper is organized as follows. In Section \ref{sec_model} we will introduce the model and define the reference mathematical setting. In Section \ref{sec_method} the control scheme is discussed in detail and we will also show how the spectral modifications trace back to actual changes in the matrix of interactions. In Section \ref{sec_2dim} the method is applied to a simple bidimensional Lotka-Volterra model, while the extension to arbitrarily large systems is discussed in Section \ref{sec_num}. Finally, in Section \ref{sec_concl} we sum up and draw our conclusions. Relevant technical material is provided in Appendices.

\section{The model}
\label{sec_model}

        We shall hereafter consider the coupled evolution of $N$ species and denote by $x_i$, $i=1, ..., N$, their associated concentrations. We will operate under the deterministic viewpoint and deliberately omit any source of stochastic disturbance, be it endogenous (demographic noise) or exogenous (external perturbation). The evolution of the ecosystem is hence described by the following set of first-order differential equations: 
		
	\begin{equation}
	\dot{x}_i=x_i(r_i-s_ix_i) + x_i\sum_{j\neq i}A_{ij}x_j\ \ \ i,j=1,...,N.
	\label{eco_eq}
	\end{equation}
	
The self-reaction term is here assumed logistic, for pedagogical reasons. We will subsequently relax this working hypothesis and generalize the analysis so as to account for an extended family of nonlinear reaction terms. In the above equations, $r_i$ stands for the intrinsic growth rate of species $i$, while $s_i$ is inversely proportional to the assigned carrying capacity. Matrix $\boldsymbol A$, in general asymmetric, defines the interactions among species. More specifically, the scalar entry $A_{ij}$ encodes the effect exerted by species $j$ on species $i$. The magnitude of $A_{ij}$ weights the strength of the interactions. The sign of $A_{ij}$, respectively $A_{ji}$, defines the specific nature of the interaction
 between species $i$ and $j$. Adopting a wording which is inspired to ecological applications: exploitation $(+,-)$, competition $(-,-)$, cooperation $(+,+)$, commensalism $(+,0)$, amensalism $(-,0)$ or null interaction $(0,0)$. The coupling among species is shaped by a quadratic term, which scales like the product of relative concentrations.  This implies assuming the interaction to be mediated by pair exchanges, as it is customarily the case in ecology.  On a more fundamental level, the coupling term here introduced will enable us to generalize the analysis reported in \cite{Cencetti_etal17} beyond standard diffusion.

  Let us denote by $\boldsymbol{x^*}$ the fixed point of the dynamics and assume for the sake of simplicity that all entries $x^*_i$ are different from zero. In formulae we have:	
	
	\begin{equation}
	r_i-s_ix_i^*+\sum_{j\neq i}A_{ij}x_j^*=0\ \ \ i=1,...,N
	\label{fix_p}
	\end{equation}	
which essentially defines the sole non trivial equilibrium eventually attained by the system under scrutiny. As alluded to in the introductory Section, we are here concerned with the resilience of the system, namely its inherent capability to recover from perturbations. Stated it differently, we shall elaborate on the conditions which make the 
fixed point stable, according to a linear stability analysis. To this end, it is convenient to introduce the rescaled variable $y_i\equiv x_i/x_i^*$. In the new variables the fixed point is homogeneous and reads 
$y_i^*=1$ $\forall i$. The dynamics of system \eqref{eco_eq} can be cast in the form:
 
	\begin{equation}
		\dot{y}_i=y_i(r_i-\tilde{s}_iy_i+\sum_{j\neq i}B_{ij}y_j)
		\label{rescaled}
	\end{equation}
where $\tilde s_i=s_ix_i^*$, $B_{ij}=A_{ij}x_j^*$ and the equation \eqref{fix_p} assumes the form $r_i-\tilde{s}_i+\sum_{j\neq i}B_{ij}=0$.

To assess the stability of the fixed point, we set $y_i=1+v_i$ and Taylor expand at the first order in the perturbation amount. This yields the following linear system for the evolution of the imposed perturbation:

	\begin{equation}
	\label{linear_eq_1}
	\dot v_i=-\tilde s_iv_i+\sum_{j\neq i}B_{ij}v_j \equiv \sum_jC _{ij}v_j.
	\end{equation}

The stability of the fixed point is ultimately controlled by the eigenvalues of the matrix $\boldsymbol{C}$ obtained by adding the elements $-\tilde{s}_i$ on the diagonal of matrix $\boldsymbol{B}$. The system is unstable when at least one eigenvalue of $\boldsymbol{C}$ has a positive real part. In the following, we will discuss a control procedure to stabilise a fixed point, that is initially engineered to be unstable. The method builds on the technique discussed in \cite{Cencetti_etal17} and aims at reshaping the coupling among interacting species. As a side observation, which will become crucial in the forthcoming analysis, we notice that the fixed point condition ($y_i^*=1$ $\forall i$) translates into a constraint for $\boldsymbol{C}$: summing the elements of $\boldsymbol{C}$ relative to row $i$ one should recover $-r_i$, i.e. $\sum_j C_{ij} = -r_i$.

	\section{Topological control scheme}
	\label{sec_method}

The proposed control strategy aims at modifying an initially unstable system of the type described above to yield an equivalent analogue which preserves the  form \eqref{eco_eq} while admitting a stable non trivial equilibrium. As it shall be argued hereafter, we can either enforce the stability of the original, assumed unstable, fixed point, or, alternatively, steer the system towards a different equilibrium. In the former case we shall also alter the original carrying capacity (a parameter which indirectly encodes for the interaction with the surrounding environment), while in the latter the envisaged protocol will solely impact the network of interspecies couplings, leaving unchanged individual reaction parameters. In both cases,  the intrinsic growth rates $r_i$, are kept unvaried. Stable ecological networks display remarkable topological characteristics, as discussed in a series of  papers devoted to this topic \cite{AllesinaTang12, CoyteSchluterFoster15}. Operating along this line, we will isolate and discuss a selected gallery of features that appear to be recurrently shared by the ecological networks stabilized as outlined in the following. 

Since, by definition, the structure of \eqref{eco_eq} is invariant under the foreseen procedure,  inspecting the linear stability of the ensuing equilibrium implies dealing with a system  of the type $\dot{v}_i=\sum_jC'_{ij}v_j$, where $\boldsymbol C'$ is obtained from $\boldsymbol C$, defined as in equation (\ref{linear_eq_1}), via the devised control algorithm. Requiring the sought stability is, in turn, equivalent to constrain the spectrum of $\boldsymbol C'$ in the left portion of the complex plane, namely to set the real part of the associated eigenvalues to negative values. Our goal, pursued hereafter, is to elaborate on a rigorous mathematical procedure, which is both anchored to first principles and potentially minimally invasive, to derive $\boldsymbol C'$ from $\boldsymbol C$. Importantly, the obtained matrix $\boldsymbol C'$ should match the condition $\sum_j C'_{ij} = -r_i$, for the homogeneous fixed point to exist in terms of the rescaled variables $y_i$ (recall that, by hypothesis, $r_i$ is  frozen to its original value). The effect of the control will be then gauged by tracing the modifications back to the underlying nonlinear framework, i.e. by evaluating the impact produced on the relevant dynamical parameters. 

As a  first step in the analysis, we write the linear equation (\ref{linear_eq_1}) in the equivalent form:

      \begin{equation}
	\label{linear_eq_2}
	\dot v_i=-r_i v_i+\sum_{j}D_{ij}v_j.
	\end{equation}
where the definition of $\boldsymbol D$ follows trivially. Recalling that $\sum_j C_{ij} = -r_i$ by virtue of the aforementioned fixed point condition, it is immediate to conclude that 
$\boldsymbol D$ is a zero-row-sum matrix, namely $\sum_j D_{ij} =0$. The next step is to diagonalize matrix $\boldsymbol D$. Formally, we set 
$\boldsymbol \Phi^{-1}\boldsymbol{ D}\boldsymbol \Phi=\boldsymbol \Lambda$, where $\boldsymbol \Phi$ is the matrix whose columns are the eigenvectors of $\boldsymbol{D}$ and $\boldsymbol \Lambda$ the diagonal matrix formed by the corresponding eigenvalues.  Diagonalizability of matrix $\boldsymbol{D}$ is hence a necessary requirement for the method to hold.
The idea is now to calculate the (minimal) shifts $\delta\Lambda^{(\alpha)}$, $\alpha=1,...,N$, to be applied to the eigenvalues of  $\Lambda^{(\alpha)}$ of matrix $\boldsymbol{D}$, for the homogeneous fixed point  $y_i^*=1$ $\forall i$ to prove linearly stable. Recall that the stability of this latter fixed point is eventually dictated by the spectrum of $\boldsymbol C$ (or, more precisely, by its controlled version $\boldsymbol{C'}$), from which the zero-row-sum counterpart $\boldsymbol D$ originates. The needed corrections $\delta\Lambda^{(\alpha)}$ are organized in a $N \times N$ diagonal matrix $\boldsymbol{\delta\Lambda}$ (with $\delta\Lambda_{kk}=\delta\Lambda^{(k)}$) to be added to matrix $\boldsymbol \Lambda$. The key point is how to choose the 
entries of  $\boldsymbol{\delta\Lambda}$ for the control to return an effective, moderately intrusive (in terms of the modifications apported on the spectrum),  stabilization. 

To answer this question, we proceed as if the original eigenvalues $\boldsymbol{\Lambda}$ were perturbed by the finite amount $\boldsymbol{\delta\Lambda}$ and recover a matrix $\boldsymbol{D'}$, which displays the modified spectrum, via the inverse transformation $\boldsymbol{D'} \equiv \boldsymbol{\Phi}(\boldsymbol{\Lambda}+\boldsymbol{\delta\Lambda})\boldsymbol{\Phi}^{-1}$. By construction $\boldsymbol{D'} $ commutes with $\boldsymbol{D}$, as the two matrices share the same set of eigenvectors. Notably, the corrections $\boldsymbol{\delta\Lambda}$ can be chosen such that matrix $\boldsymbol{D'} $ is also zero-row-sum, as $\boldsymbol{D}$ is. This is rigorously proven in Appendix \ref{app_real_stoch}, building on the derivation reported in \cite{Cencetti_etal17}. Moreover,  $\boldsymbol{D'}$ is real: this property is also inherited from $\boldsymbol{D}$, as shown again in Appendix \ref{app_real_stoch}. We are thus brought back to the linear problem:

      \begin{equation}
	\dot v_i=-r_iv_i+\sum_j D'_{ij}v_j
	\end{equation}
or, equivalently to $\dot v_i=\sum_jC'_{ij}v_j$, where:
	\begin{equation}
	C'_{ij}= \left\{\begin {array}{ll} 
	 D'_{ij}\ \ \ \ \ \ \text{if}\ i\neq j\\
	 D'_{ii}-r_i\ \ \text{if}\ i=j.\\
	\end{array}\right.
	\label{C'}
	\end{equation}

\begin{figure}
\captionsetup{justification=raggedright,singlelinecheck=false}
\includegraphics[width=10cm]{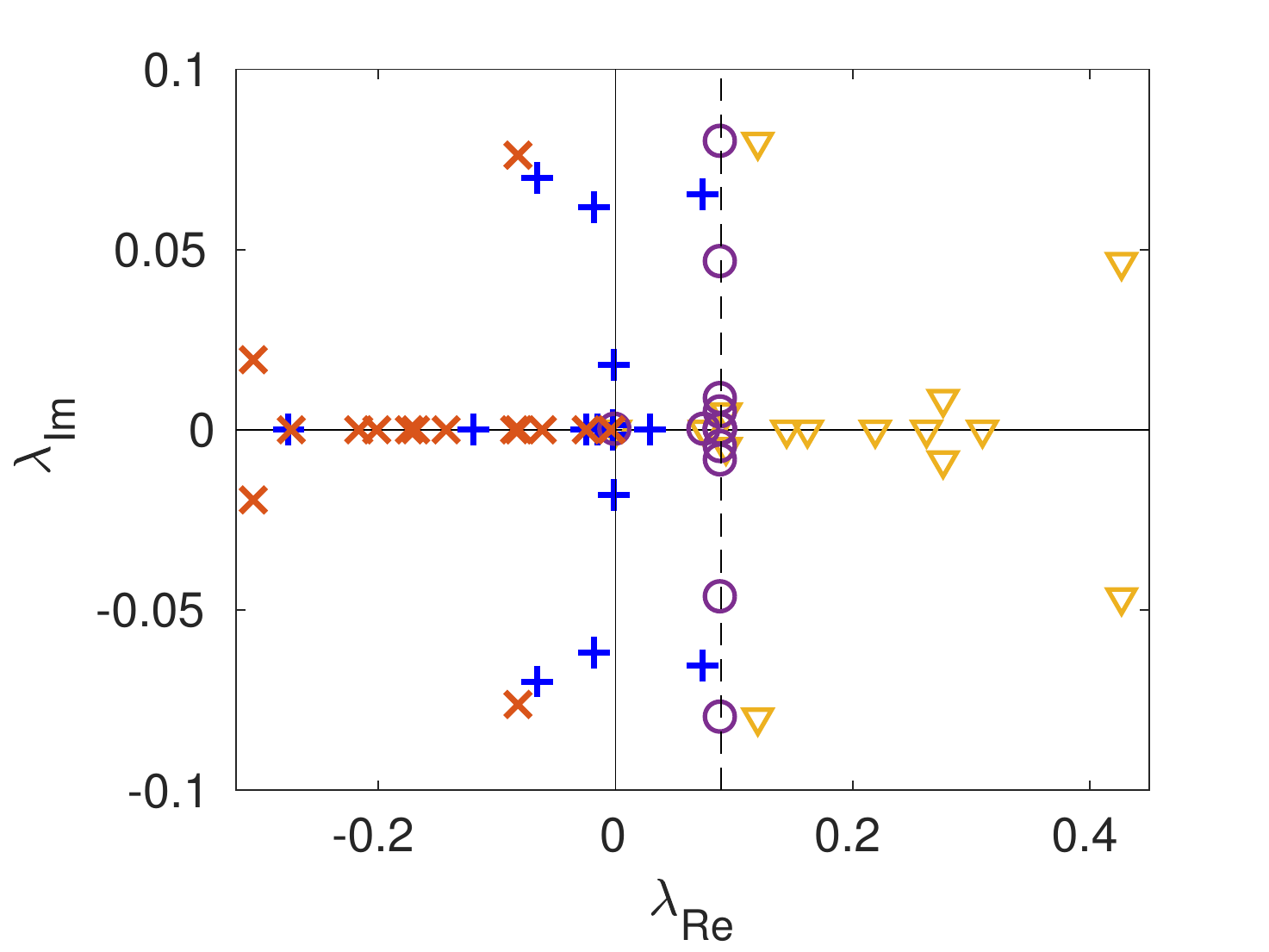}
\caption{The eigenvalues of $\boldsymbol{C}$, (blue) plus symbols, signals an initial instability. This is also seen at the level of the spectrum of $\boldsymbol{ D}$, (yellow) triangles, which partially extends beyond the critical vertical (dashed) line, located at $r_{min}$. Circles (purple) and crosses (red) refer respectively to the eigenvalues of $\boldsymbol{ D'}$ and $\boldsymbol{C'}$, pointing to the recovered stability.}
\label{0_stability_withB_N15}
\end{figure}

By construction $\sum_j C'_{ij} = -r_i$, since  $\sum_j D'_{ij} = \sum_j D_{ij}  = 0$. This is a crucial observation which makes it possible to interpret $\boldsymbol{C}'$ as the Jacobian matrix 
associated to a rescaled nonlinear problem of the type (\ref{rescaled}) where parameters $\boldsymbol r$ remain unchanged. 	The zero-row-sum-property of matrix $\boldsymbol{D'}$ is not a necessary condition to obtain stability but represents a useful requirement to help interpreting the results in terms of the original variables.\\ 
We now return to discussing the selection of the elements of the shift matrix $\boldsymbol{\delta\Lambda}$. These latter are to be chosen so as to constrain the spectrum of $\boldsymbol{C}'$ to the left hand side of the imaginary plane, thus ensuing the desired stability.  
From \eqref{C'}, it is clear that the eigenvalues of $\boldsymbol{D}'$ are positioned, in the complex plane, on the right of  those stemming from matrix $\boldsymbol{C}'$. The relative separation between the two respective spectra can be somehow quantified through $\boldsymbol{r}$. To make this observation rigorous, we recall the celebrated Gershgorin theorem \cite{Bell65}:
the eigenvalues of a given matrix are included in disks defined by the elements of the matrix itself. More specifically, the $i$-th Gershgorin disk of matrix $\boldsymbol{D}'$ corresponds to the relative disk of matrix $\boldsymbol{C}'$, translated to the right by the scalar quantity $r_i$. Unfortunately, it is not trivial to relate the index $i$ (running on the nodes) to the eigenvalues (sorted with the index $\alpha$).  To enforce stability, and assuming the worst case scenario, we shall assign the (real) shifts $\delta\Lambda^{(\alpha)}$ so that
all eigenvalues of $\boldsymbol{D}'$ have their real part smaller than $r_{min}$, the minimum of all the entries of vector $\boldsymbol{r}$. In practical terms, the imposed corrections $\delta\Lambda^{(\alpha)}$ are chosen as:

	\begin{equation}
	\delta\Lambda^{(\alpha)}= \left\{\begin {array}{ll} 
	R - \Lambda^{(\alpha)}_{Re}\ \ \ \text{if}\ \Lambda^{(\alpha)}_{Re}>r_{min}\\
	0\hspace{2cm} \text{otherwise}\\
	\end{array}\right.
	\label{deltaL}
	\end{equation}
where the scalar quantity $R$ has been introduced such that $R<r_{min}$. To help visualizing the whole procedure we report an illustrative example in Figure \ref{0_stability_withB_N15}, without insisting on the specific selection of the involved parameters. The eigenvalues of the original matrix $\boldsymbol{C}$ are displayed in the complex plane with (blue) plus symbols: the homogeneous fixed point of the rescaled equations (\ref{rescaled}) is therefore unstable, as the spectrum protrudes in the right half-plane. The eigenvalues of  $\boldsymbol{D}$ are shown with (yellow) triangles and extends on the right of the vertical dashed line, which is traced at $r_{min}$. The (purple) circles stands for the eigenvalues of the controlled matrix $\boldsymbol{D}'$: as anticipated, they are confined on the left of the vertical dashed line, the closer to the line the less invasive the control imposed on the population of unstable modes.  Finally, and as predicted, the spectrum of  the controlled Jacobian $\boldsymbol{C}'$ is contained in the negative half-plane, thus implying asymptotic stability.

As it should be clear from the above,  the control protocol assumes dealing with constant  $r_i$ parameters. Given this constraint, two viable strategies are envisaged to re-parametrize the original system, in light of the outcome of the control scheme. We have in fact 

    		\begin{equation}
		 D'_{ij}= \left\{\begin {array}{ll} 
		r_i-s_i'{x_i^*}'\hspace{7mm} \text{if}\ i=j\\
		B'_{ij} = A'_{ij}{x_j^*}'\hspace{2mm} \text{if}\ i\neq j\\
		\end{array}\right.
		\label{bivio}
		\end{equation} 
which allows in principle to define the new coupling strengths, as encoded in $\boldsymbol A'$, the novel fixed point ${\boldsymbol x^*}'$ and the modified inverse carrying capacities $\boldsymbol s'$. A first strategy to finalize the transformation suggests leaving the parameters $\boldsymbol s$ unchanged, namely $\boldsymbol s'=\boldsymbol s$. In practical terms, we assume that the reaction parameters, which characterize the dynamics of each species when evolved on a isolated patch, remain unchanged. The ecosystem can achieve stabilization, by just reshaping the underlying networks of interlaced dependencies. From equation \eqref{bivio}, we have therefore:

	\begin{equation}
	\begin {array}{ll} 
	{x_i^*}'=\frac{r_i-D'_{ii}}{s_i}\\
	A'_{ij}=\frac{D'_{ij}}{{x_j^*}'}=\frac{D'_{ij}s_j}{r_j-D'_{jj}}.\\
	\end{array} 
	\label{0_A'x'}
	\end{equation}
	
Setting paired interactions as specified by matrix $\boldsymbol A'$  guarantees the stability of the associated, and consistently modified, fixed point $\boldsymbol{{x^*}'}$.  The quantities ${x_i^*}'$ should be positive defined (at least when ecological applications are concerned), which in turn translates into the additional requirement  
\begin{equation}
	r_i-D'_{ii}>0\ \forall i.
	\label{applicability}
\end{equation}
	
The second strategy consists of modifying the parameters $\boldsymbol s$, together with the matrix $\boldsymbol A$, leaving unchanged the fixed point $\boldsymbol x^*$. The carrying capacity $\boldsymbol s$ is prone to environmental influences, and, as such, it can be imagined to be tunable with some degree of realism. This is opposed to the growth parameter $\boldsymbol r$,  constrained, among other factors, by  species genetics, and thus assumed constant throughout the procedure. From equation \eqref{bivio}:
	\begin{equation}
	\begin {array}{ll} 
	s_i'=\frac{r_i-D'_{ii}}{x^*_i}\\
	A'_{ij}=\frac{D'_{ij}}{x_j^*}.
	\end{array} 
	\end{equation} 
	The additional condition  $s_i'>0$ should be imposed, which again amounts to requiring  equation \eqref{applicability} to hold. In other words, condition \eqref{applicability} is  a general    constraint that the control scheme is bound to verify, for the specific ecologically inspired application, here discussed. We shall refer to condition \eqref{applicability} as to the \textit{applicability constraint} and elaborate on its implications hereafter. The idea is to find a suitable value for $R$ in order to match the condition \eqref{applicability}, and, hence, to make the control scheme applicable. To begin, let us assume that we are allowed to modify all eigenvalues of the considered spectrum, and not just the limited sub-set that triggers the system unstable.  Then, it is enough to impose $\delta\Lambda^{(i)}\leq\tilde{s}_{min}$ $\forall i$, which is in principle always possible, in order to automatically verify \eqref{applicability}. The system is therefore always controllable when all eigenvalues are to be modified.\\

Consider now the more interesting case where a subset $\mathcal{M}$ of elements of the whole spectrum is the target of the control. One has to face the following restrictions:
    \begin{itemize}
        \item if an index $i$ exists such that $\sum_{j\in \mathcal M}\Phi_{ij}\Phi_{ji}^{-1}<0$ and $\tilde s_i+ \sum_{j\in \mathcal M}\Phi_{ij}\Lambda_{Re}^{(j)}\Phi_{ji}^{-1} - r_{min}\sum_{j\in \mathcal M}\Phi_{ij}\Phi_{ji}^{-1}<0$, then constraint \eqref{applicability} is never matched and the system is not controllable with the  above discussed technique.
        
        \item if an index $i$ exists such that $\sum_{j\in \mathcal M}\Phi_{ij}\Phi_{ji}^{-1}>0$ and $\tilde s_i+ \sum_{j\in \mathcal M}\Phi_{ij}\Lambda_{Re}^{(j)}\Phi_{ji}^{-1} - r_{min}\sum_{j\in \mathcal M}\Phi_{ij}\Phi_{ji}^{-1}<0$ (suppose node indices are sorted so that such an index $i$ is in the subset of nodes $1,...,\tilde n$ with $\tilde n < N$), than the applicability condition \eqref{applicability} is verified only if the following statement holds true:
        \begin{equation}
            \max_{i\in[1,\tilde n]} \biggl(r_{min} - \frac{\tilde s_i+ \sum_{j\in \mathcal M}\Phi_{ij}\Lambda_{Re}^{(j)}\Phi_{ji}^{-1}}{\sum_{j\in \mathcal M}\Phi_{ij}\Phi_{ji}^{-1}}\biggr)
            \leq 
            \min_{i\in[n+1,N]} \biggl(r_{min} - \frac{\tilde s_i+ \sum_{j\in \mathcal M}\Phi_{ij}\Lambda_{Re}^{(j)}\Phi_{ji}^{-1}}{\sum_{j\in \mathcal M}\Phi_{ij}\Phi_{ji}^{-1}}\biggr)
        \end{equation}
        where $[n+1,N]$ numbers the set of indices $i$ for which $\sum_{j\in \mathcal M}\Phi_{ij}\Phi_{ji}^{-1}<0$.
    \end{itemize}

The above conditions (derived in Appendix \ref{app_controllability}) are to be carefully checked before attempting to control the system under scrutiny via the procedure that we have here illustrated and which is ultimately aimed at recalibrating the weights of the underlying couplings.
	
\begin{figure}
\captionsetup{justification=raggedright,singlelinecheck=false}
\includegraphics[width=12cm]{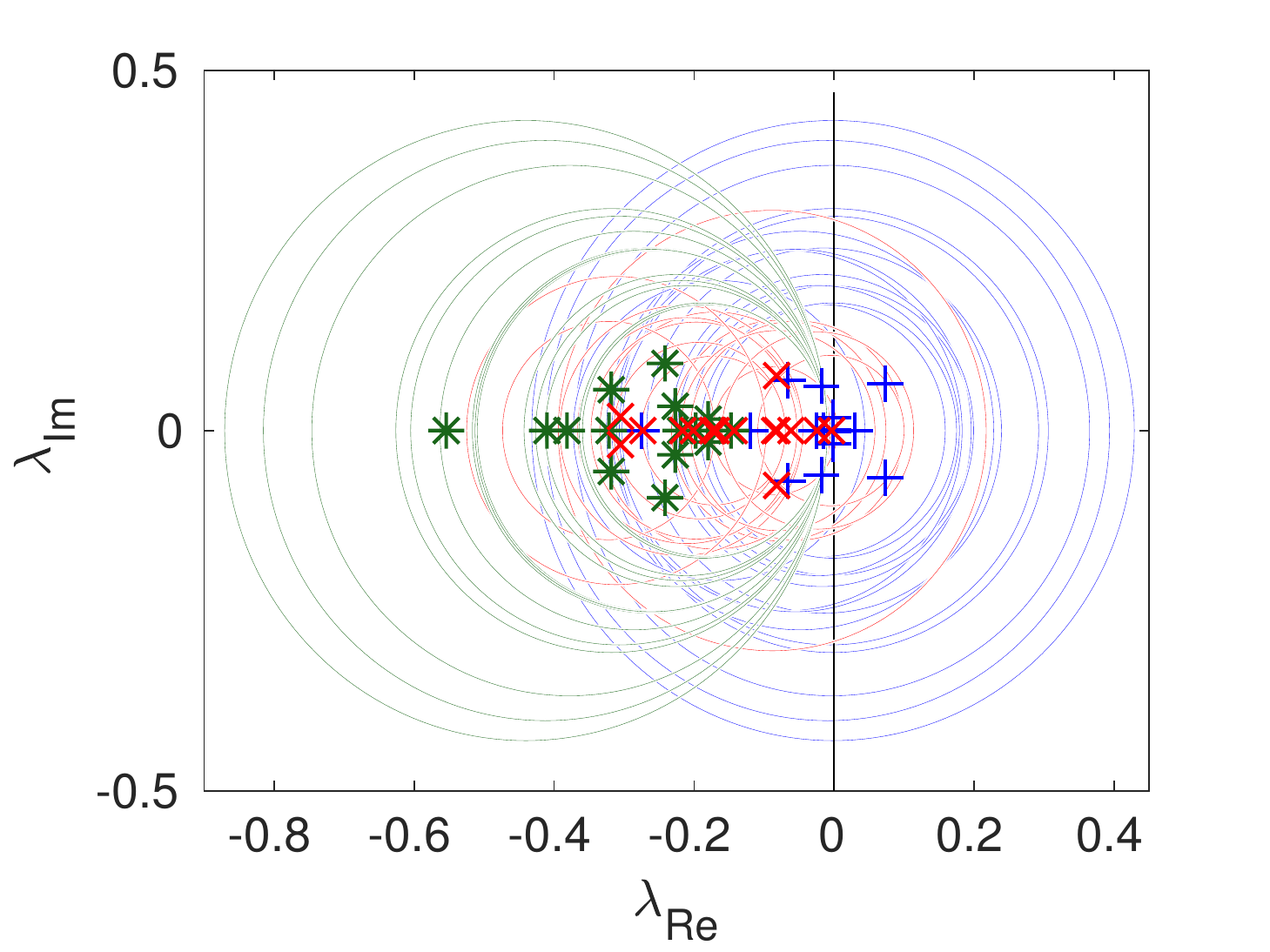} 
\caption{Eigenvalues in the complex plane for a particular choice of matrix $\boldsymbol{C}$ (blue plus symbols): the system is unstable as the symbols invade the region with positive $\lambda_{Re}$. The (blue) circles represent the associated Gershgorin disks. Red symbols stand for the eigenvalues of matrix $\boldsymbol{C'}$, which fall in the negative part of the complex plane, while their corresponding circles protrude in the positive half-plane. Green symbols refer to the alternative control scheme \eqref{Gersh_cond}: now the Gershgorin circles are  contained in the negative half-plane. The spectrum obtained by following this latter route to stability is well inside the region of stability, at variance with that generated by the former approach, which sits at the border of stability. In this respect, the method that we have depicted, and which necessitates reshaping the underlying network of contacts, can be thought as minimally invasive. }
\label{fig_Gersh}
\end{figure}

	
Before concluding this Section, we briefly mention an alternative control strategy which builds on the already mentioned  Gershgorin theorem. In the above analysis we have reshaped the networks of contacts and altered either the fixed point $\boldsymbol{x^*}$ or the carrying capacity $s_i$, while preserving the values of the growth factors $r_i$. The alternative route to stabilization that we shall hereafter discuss follows a dual path: the only parameters to be tuned are the growth rates $r_i$. Recalling that $\boldsymbol C'$ is real (from definition of \eqref{C'}, being $\boldsymbol{ D}'$ real), it is clear that each eigenvalue has its real part smaller than $C'_{ii}+\sum_{j\neq i}|C'_{ij}|$, which is the rigthmost point in the complex plane of the $i$-th Gershgorin disk for matrix $\boldsymbol C'$. To enforce stability, we could then require that all the Gershgorin circles are included in the left half-plane:
	\begin{equation}
	C'_{ii}+\sum_{j\neq i}|C'_{ij}|\leq0 \Rightarrow D'_{ii}-r_i+\sum_{j\neq i}|D'_{ij|}\leq0 \Rightarrow D'_{ii}+\sum_{j\neq i}|D'_{ij|}\leq r_i
	\label{Gersh_cond}
	\end{equation}
	which implies that the Gershgorin disks computed for matrix $\boldsymbol{D'}$ should be contained in the semi-plane constrained, from the right, by the  vertical line  located at  $r_i$ for any selected $i$, or, better, at the left of $r_{min}$, for all $i$. The shift of the Gershgorin disks can be performed by modifying the diagonal of matrix $\boldsymbol C$, which is equivalent to changing the vector of parameters $\boldsymbol{r}$. This alternative control strategy proves however more invasive in terms of the perturbation that is produced on the original spectrum. This is clearly testified in Figure \ref{fig_Gersh}, where the original (unstable by construction) spectrum is compared to those obtained applying the two control strategies outlined above.

In the next Section we will begin by applying the developed method to a simple example, where just two species are made to evolve. This application bears pedagogical interest and it will pave the way to inspecting the general setting, on which we shall report in the subsequent Section.

	\section{Stabilizing the Lotka-Volterra dynamics}
	\label{sec_2dim}
	
\begin{figure}
		\captionsetup{justification=raggedright,singlelinecheck=false}
		\subfigure[]{
			\includegraphics[width=8cm]{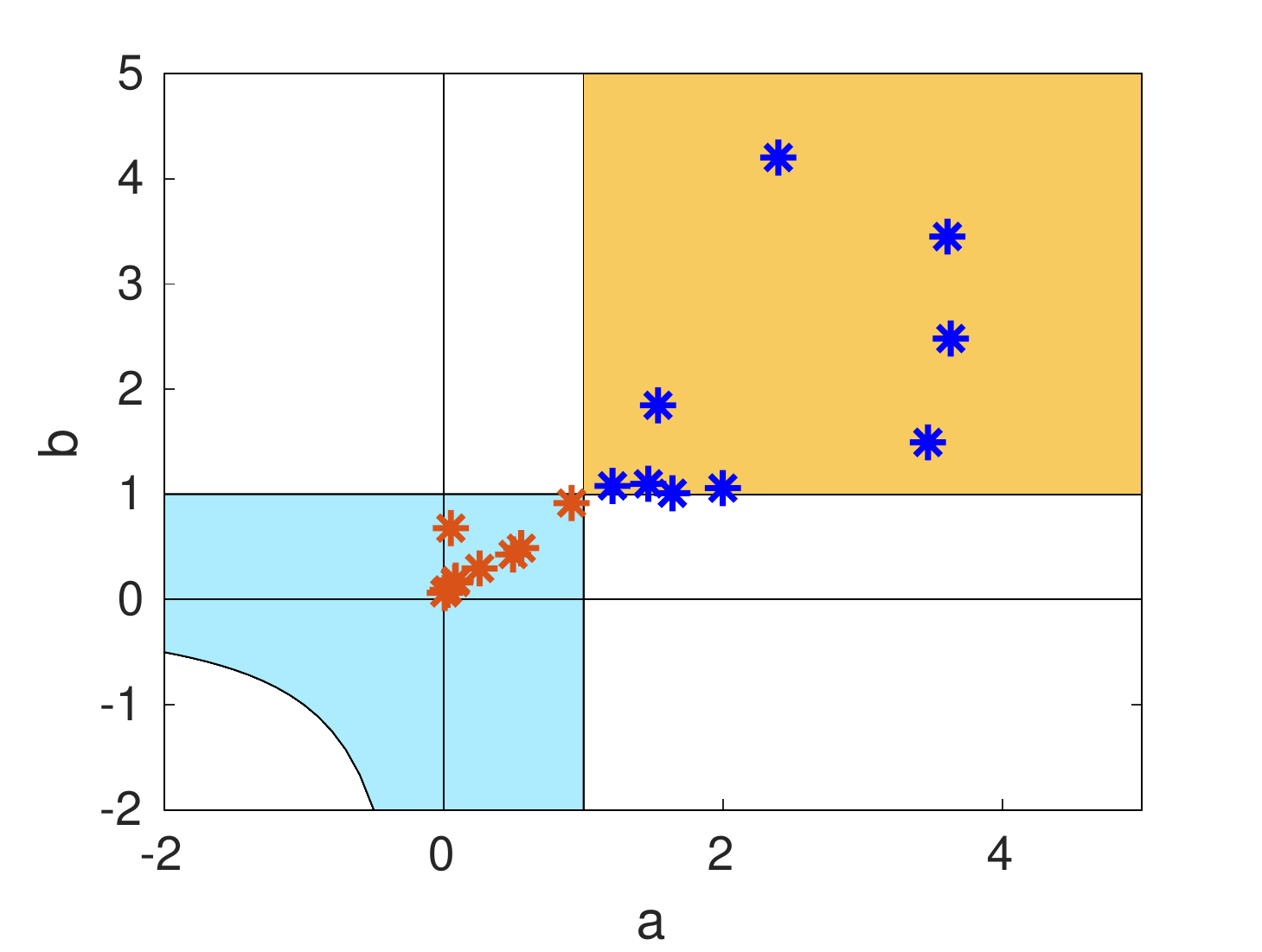}
		}
		\subfigure[]{
			\includegraphics[width=8cm]{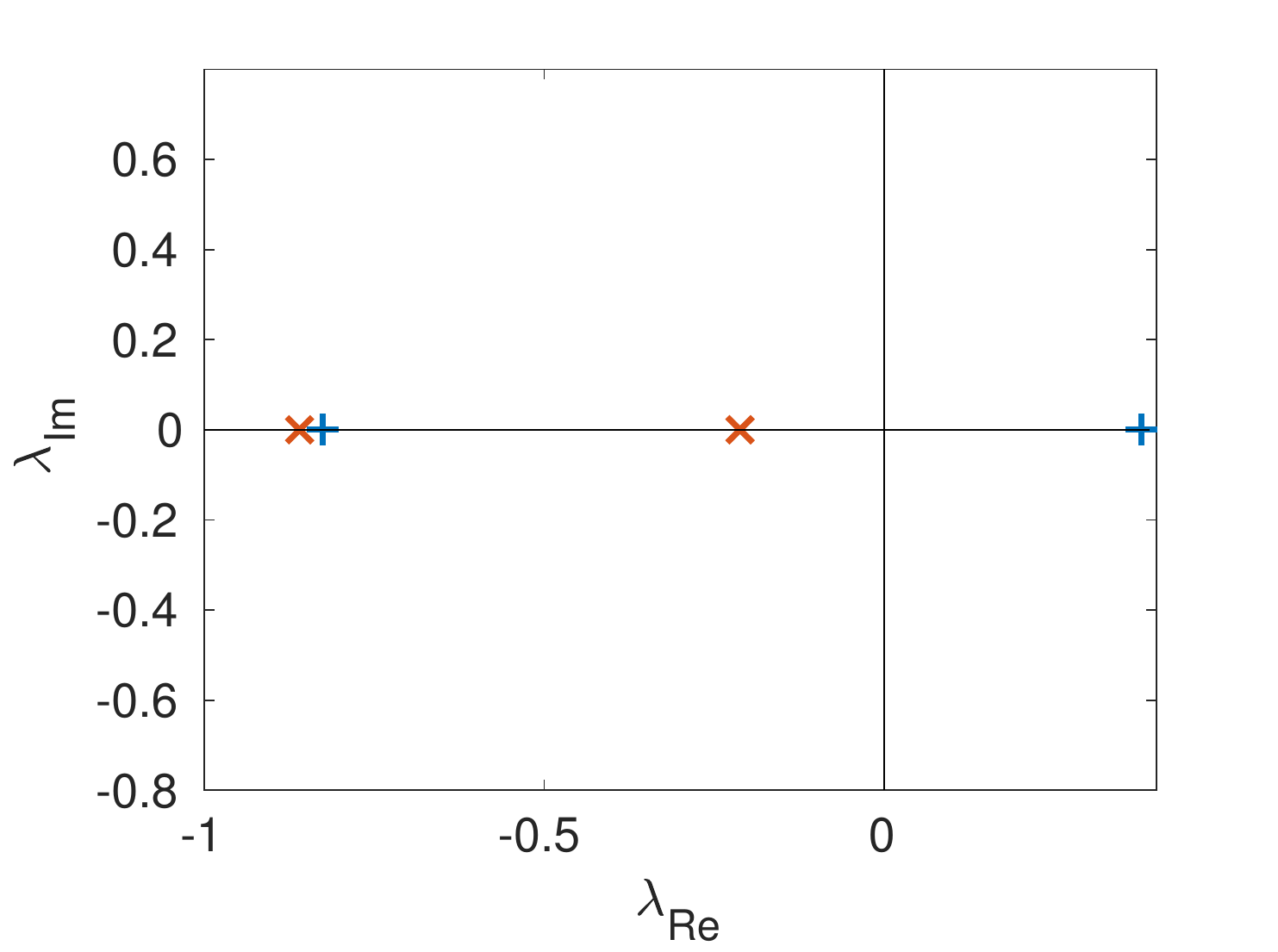}
		}		
		\caption{Panel (a): Existence and stability of the (non trivial) fixed point $(u_1^*,u_2^*)$, as a function of the parameters $a$ and $b$. In particular, the fixed point does not exist in the empty (no shading) areas (negative for $a>1 \cup b<1$ and $a<1 \cup b>1$, infinite for $a<0 \cup b<0 \cup b<1/a$); it is unstable in the rectangular (orange) shaded region, entirely contained in the first quadrant; it is stable in the other (cyan) colored region, which extends in all quadrants. The blue stars refer to the initial choice of the parameters: the system is hence unstable. After  the control procedure is applied, one obtains the red stars, which are distributed inside the region associated to linear stability. These latter symbols cluster in a limited portion of the plane, close to the threshold of instability. In this respect, the control method is minimal also in terms of the modification induced at the level of  the dynamical parameters (and not only in relation to its effects in the complex plane where the spectrum of the Jacobian is depicted).   Panel (b): original and modified spectrum are displayed for one representative case study among those depicted in panel (a). }
		\label{fig_phase_space}
	\end{figure}

With the goal of gaining further insight on the control scheme developed above, we will here consider a simple setting where just two species are made to mutually interact. Hence, we will consider hereafter  $N=2$ and label with $x_1$ and $x_2$ the mean field concentrations of the interacting species. Equations \eqref{eco_eq} reduce therefore to:

	\begin{equation}
	\left\{\begin {array}{ll} 
	\dot x_1=x_1(r_1-s_1x_1+A_{12}x_2)\\
	\dot x_2=x_2(r_2-s_2x_2+A_{21}x_1)\\
	\end{array}\right.
	\label{eq_2dim}
	\end{equation}
a framework known in the literature as the Lotka-Volterra competition model. To proceed in the analysis we define $u_i\equiv (s_i / r_i ) x_i$, $\tau\equiv r_1t$,  and eventually get the following governing equation in the rescaled variables $u_1$ and $u_2$:	
		
		\begin{equation}
	\left\{\begin {array}{lll} 
	\frac{du_1}{d\tau}=u_1(1-u_1+au_2)\equiv f_1(u_1,u_2)\\
	\\
	\frac{du_2}{d\tau}=\rho u_2(1-u_2+bu_1)\equiv f_2(u_1,u_2)\\
	\end{array}\right.
	\end{equation}
where $\rho=r_2/r_1$, $a\equiv-A_{12}\frac{r_2}{r_1s_2}$ and $b\equiv-A_{21}\frac{r_1}{r_2s_1}$. It is straightforward to prove that this system admits four equilibria: $(u_1, u_2)=(0,0)$, $(u_1, u_2)=(0,1)$, $(u_1, u_2)=(1,0)$ and $(u_1, u_2)=(\frac{1-a}{1-ab},\frac{1-b}{1-ab})\equiv(u_1^*, u_2^*)$. We shall hereafter refer to the latter equilibrium, the only one to guarantee non trivial asymptotic concentrations for both species. Notice that this is admissible only if $u_1^*$ and $u_2^*$ are positive and finite, which, in turn, implies that the parameter space $(a,b)$ is restricted to the colored region of Figure \ref{fig_phase_space}, i.e.
	
	\begin{itemize}
		\item $a>1$, $b>1$
		\item $a<1$, $b<1$, $b<1/a$
	\end{itemize}

	The stability of the selected fixed point is determined by the  Jacobian matrix
	\begin{equation}
	J_{(u_1^*,u_2^*)}=
	\frac{1}{1-ab}
	\left (
	\begin{matrix}
	a-1 & a(a-1)\\
	\rho b(b-1) & \rho(b-1)\\
	\end{matrix}
	\right )
	\end{equation}
	whose eigenvalues are 
	\begin{equation}
	\lambda_{\pm}=\frac{1}{2(1-ab)}\biggl[(b-1)(1+\rho)\pm\sqrt{(a-1)^2(1+\rho)^2-4\rho(1-ab)(a-1)(b-1)}\biggr].
	\end{equation}
	The sign of $\lambda_{\pm}$ implies that the fixed point $(u_1^*,u_2^*)$ is unstable for $a>1$ and $b>1$ and stable in the complementary domain, as depicted in Figure \ref{fig_phase_space}.\\
	
	
Working in this simplified setting, it is therefore straightforward to implement, and graphically illustrate, the stabilization protocol, as addressed in the preceding Section. 
Starting from an unstable system corresponds to setting the parameters $a$ and $b$ in the (orange) shaded sub-portion of the first quadrant of Figure \ref{fig_phase_space}. When implementing the control, panel (b) of Figure \ref{fig_phase_space},  the rescaled parameters $(a,b)$ are consequently moved to the other (cyan) shaded domain displayed in Figure \ref{fig_phase_space} (a), i.e. the region deputed to stability. Different symbols, as depicted in the Figure \ref{fig_phase_space} (a), refers to distinct choices of the initial model parameters. In all cases, the stabilization is successfully produced and, more importantly, the modified parameters $(a,b)$ tend to cluster in a limited portion of the stability domain, close to the boundary of instability. This observation  suggests again that the devised control acts  by producing a  somehow minimal perturbation to the original model.  Following the alternative control recipe based on the Gershgorin theorem 
produces much more invasive changes (here not shown).

The controlled matrix of interaction, as well as the novel set of dynamical parameters, can be readily obtained from the modified quantities $a'$ and $b'$,  following one of the interpretative scenarios discussed with reference to the $N$-dimensional case.  Both strategies require altering the matrix of couplings $\boldsymbol{A}$ to eventually obtain its modified counterpart here denoted with $\boldsymbol{A'}$. 
More specifically:

	\begin{itemize}
		\item Changing the fixed point $\boldsymbol x^*$.  The 
		 parameters $\boldsymbol r$ and $\boldsymbol s$ stay unchanged. Hence:

		\begin{eqnarray}
		A'_{12} &=& -a' \frac{r_1 s_2}{r_2}\\
		A'_{21}&=&-b' \frac{r_2 s_1}{r_1}\\		
		\end{eqnarray}

		The new fixed point $(x_1^*)'$ , $(x_2^*)'$ is obtained by solving the self-consistent equations:
		\begin{eqnarray}
		r_1-s_1(x_1^*)'+A'_{12}(x_2^*)' =0\\
		r_2-s_2(x_2^*)'+A'_{21}(x_1^*)' =0.\\
		\end{eqnarray}
		
		\item Modifying the parameters $\boldsymbol s$. The fixed point and $\boldsymbol r$ are not varied. One gets:
		
		\begin{equation}
		\left\{\begin {array}{ll} 
		a'=-A'_{12}\frac{r_2}{r_1s_2'}\\
		b'=-A'_{21}\frac{r_1}{r_2s_1'}\\
		\end{array}\right. 
		\end{equation}
		which, together with the fixed point condition:
		\begin{equation}
		\left\{\begin {array}{ll} 
		r_1-s_1'x_1^*+A'_{12}x_2^*=0\\
		r_2-s_2'x_2^*+A'_{21}x_1^*=0.\\
		\end{array}\right.
		\end{equation}
		allows one to obtain the entries of the matrix $A'$ and the controlled parameters $\boldsymbol s$.
	\end{itemize}
	
In the first case we have to make sure that the new fixed point is admissible, which amounts to meeting the conditions $(x_1^*)'>0$ and $(x_2^*)'>0$. In the second case, one has to impose $s_1'>0$ and $s_2'>0$. It is straightforward to prove that these constraints are always satisfied when $a'$ and $b'$ fall, as they do by definition, in the stability region ($1-a'>0$, $1-b'>0$ and $1-a'b'>0$). In the next Section we move on to considering the general setting by working with an arbitrarily large ecological system consisting of $N$ interacting populations.

	\section{The general case}
	\label{sec_num}
	
	Let us now consider the general setting where an arbitrarily large number of  species is made to interact. The initial coupling strengths which exemplify the interaction among distinct populations, as encoded in the elements of matrix $\boldsymbol{A}$, are initially assigned in such a way that a non trivial fixed point exists and is linearly unstable. This is achieved by generating the random entries $A_{ij}$, according to a specific distribution that we discussed in the Appendix and exploiting again the Gershgorin theorem. In the following, we will set $N=100$ and illustrate the results which are obtained when allowing for the fixed point to be modified by the control procedure. The alternative control strategy, which leaves the original fixed point unchanged, is also analyzed and the results reported in the Appendix. As an interesting outcome, we will show that predator-prey interactions exert a stabilizing effect, as already pointed out in \cite{AllesinaTang12}. 
	
\begin{figure}
		\captionsetup{justification=raggedright,singlelinecheck=false}
		\includegraphics[width=10cm]{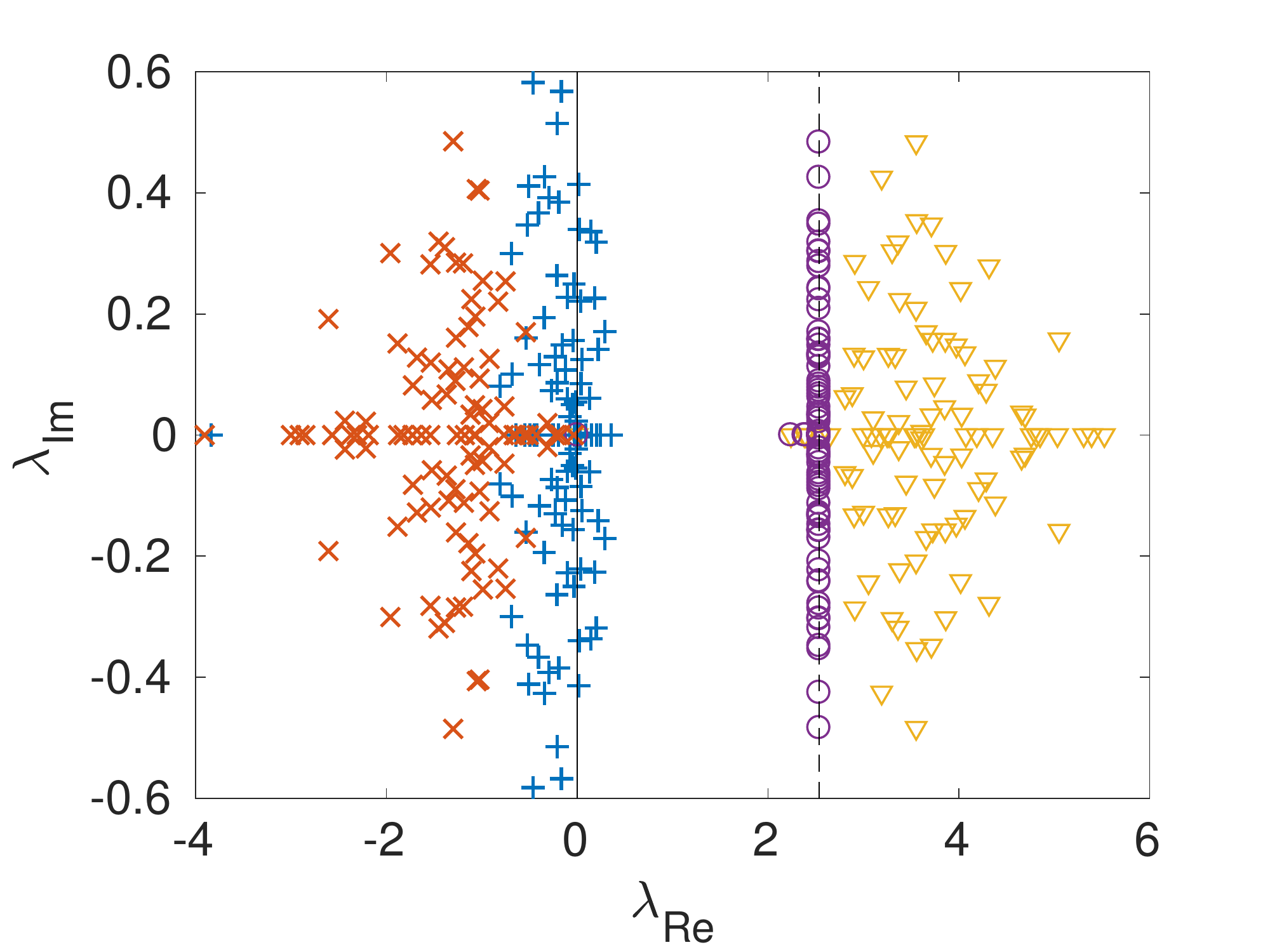}
		\caption{Eigenvalues in the complex plane, before, along and after the control procedure. The symbols are assigned as described in the caption of Figure \ref{0_stability_withB_N15}. The elements $A_{ij}$ are assigned following the scheme discussed in the Appendix. Here, $N=100$.}
		\label{0_stability_withB}
	\end{figure}

	In Figure \ref{0_stability_withB}, the original and modified spectrum are displayed in the complex plane, for one representative realization of matrix $\boldsymbol{A}$. The symbols are chosen following the same convention adopted in Figure \ref{0_stability_withB_N15}. The stabilization produced by the control is clearly demonstrated.

\begin{figure}
		\captionsetup{justification=raggedright,singlelinecheck=false}
		\subfigure[]{
			\includegraphics[width=8cm]{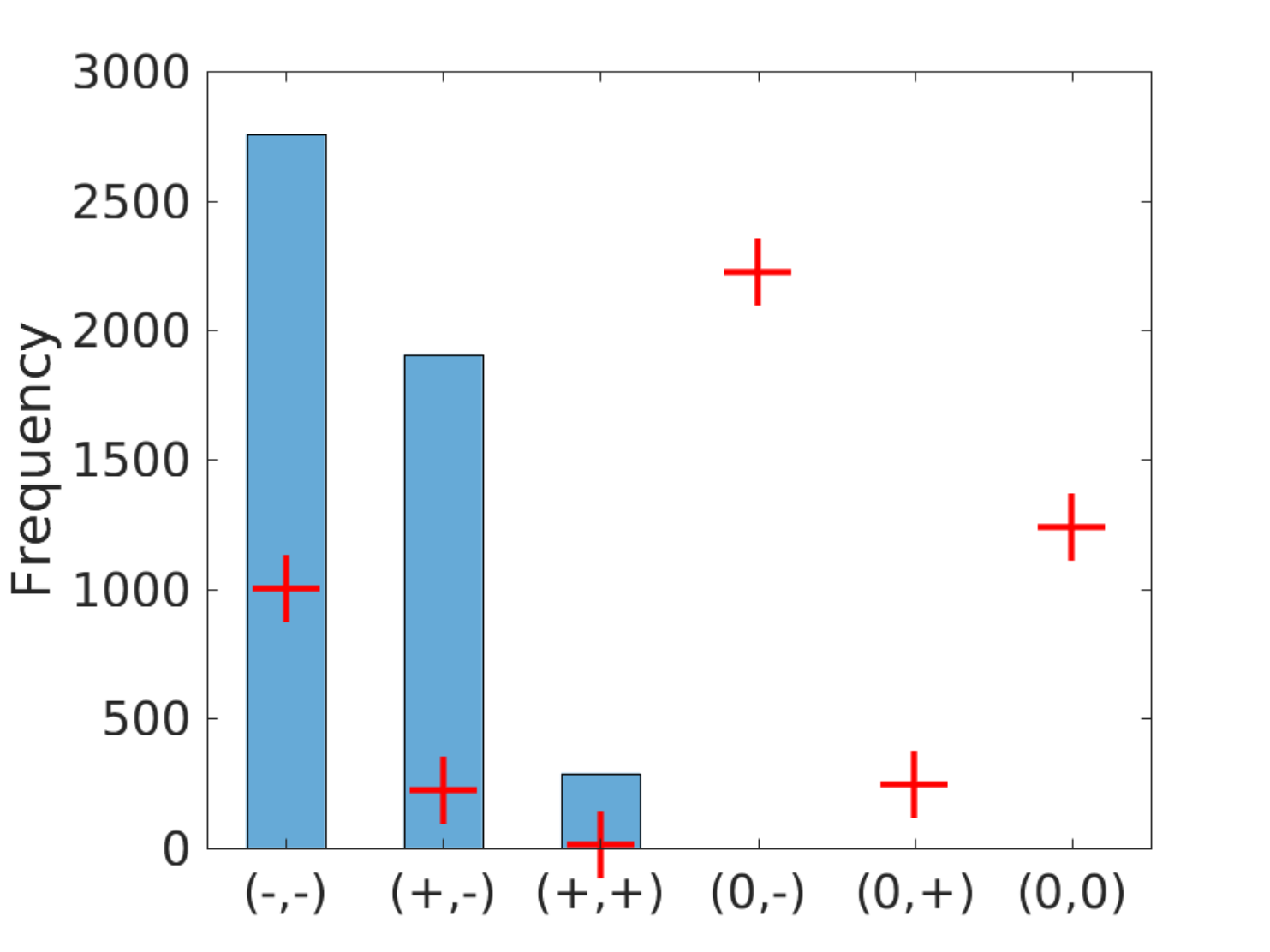}
			\label{0_histo_after}
		}
		\subfigure[]{
			\includegraphics[width=8cm]{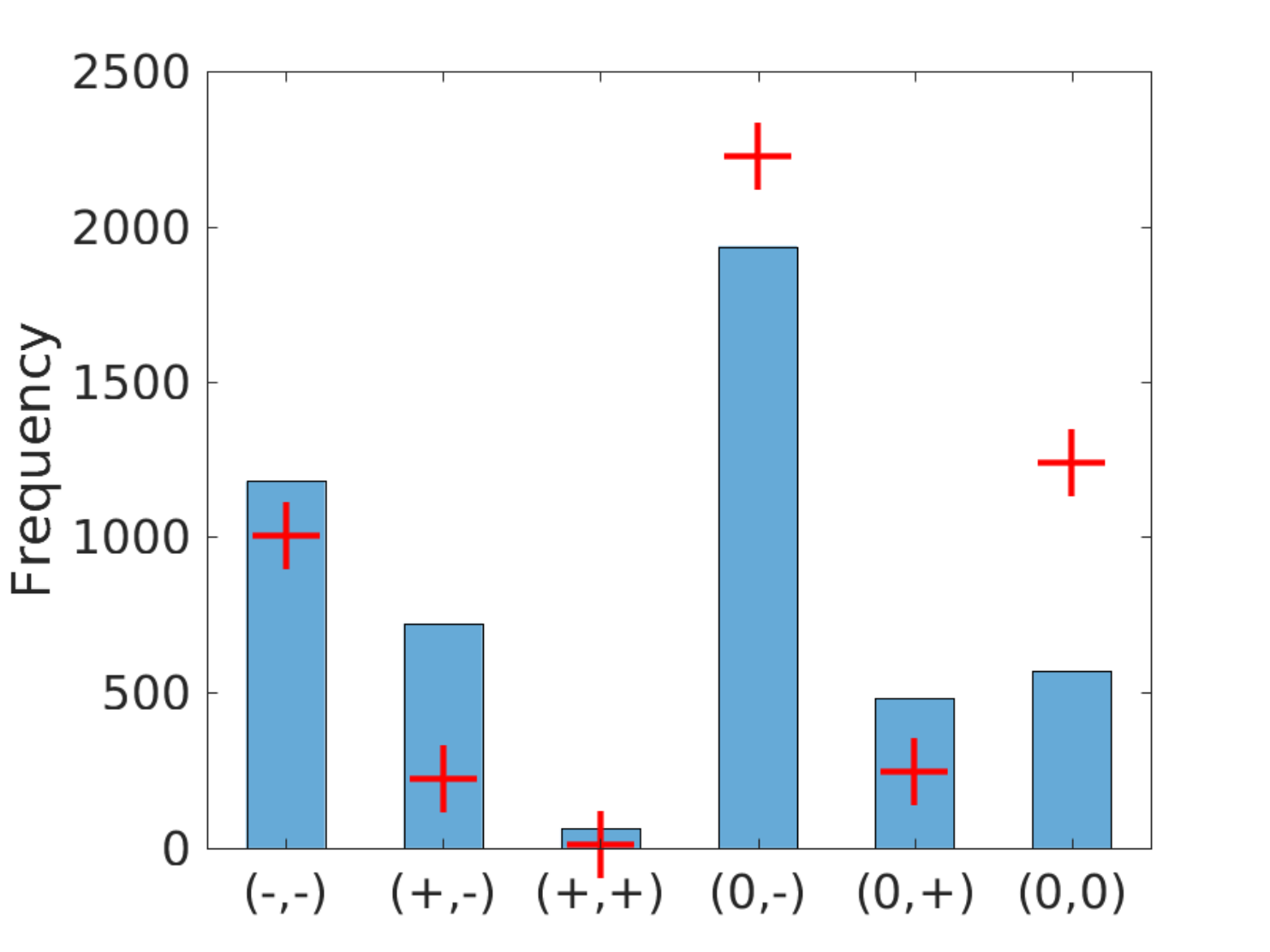}
			\label{0_histo_after_cutoff}
		}
		
		\caption{Abundances of different types of couplings between species.  (Red) plus symbols refer to the uncontrolled matrix of interaction (analytically computed in Appendix \ref{app_orig}).  
			Panel (a): the (blue) bars report on the relative abundances of different classes, as obtained after the control has been applied. Results refer to just one realization of the process. Quantitatively similar conclusions are obtained for different realizations and/or averaging over a large ensemble of them. Panel (b): the frequency of appearance of different classes is plotted after a cut-off has been applied (see main text). Here the cut off is set to $0.005$.}
		\label{ecology_histo}
	\end{figure}

	To help visualizing how the control shapes the examined ecosystem, we extract from the matrix $\boldsymbol{A}'$ the number of pairs that belong to the different classes, categorized in five six large classes, here recalled for the sake of completeness: exploitation $(+,-)$, competition $(-,-)$, cooperation $(+,+)$, commensalism $(+,0)$, amensalism $(-,0)$ or null interaction $(0,0)$. 
	The frequency of appearance of different classes  is investigated in Figure \ref{ecology_histo}. The red plus symbols point to the uncontrolled setting, and reflect the specific rule chosen for generating matrix $\boldsymbol{A}$ and its associated, unstable fixed point (see Appendix \ref{app_orig}). For the controlled matrix, only three bins are populated, see Figure \ref{ecology_histo} (a): interaction modalities that envisage a one directional coupling, or stated differently, a zero entry in matrix $\boldsymbol{A}$, are absent in the controlled scenario. The topological control activates in fact all pairwise connections, albeit often by a tiny amount.  It is then interesting to silence, a posteriori, in the controlled adjacency matrix $\boldsymbol{A}'$,  the links that are associated to  a weight (in absolute value) smaller than a given cutoff. For a sufficiently small cutoff the stability of the controlled system is preserved: the number of newly added links
	can be hence considerably reduced, by eradicating from the collection those that bear no relevance in light of the modest exerted coupling. The effect of the cut off is visible in Figure \ref{ecology_histo} (b): the final distribution of pairs resembles very closely the one generated at the beginning. Remarkably, the number of predator-prey interactions grows at the 
	detriment of the last column of the histogram, implying that the new interactions that are to be established for stability to hold belong to this class, in qualitative agreement with the analysis by Allesina et al. \cite{AllesinaTang12}.

	\

\begin{figure}
		\captionsetup{justification=raggedright,singlelinecheck=false}
		\subfigure[]
		{\includegraphics[width=10cm]{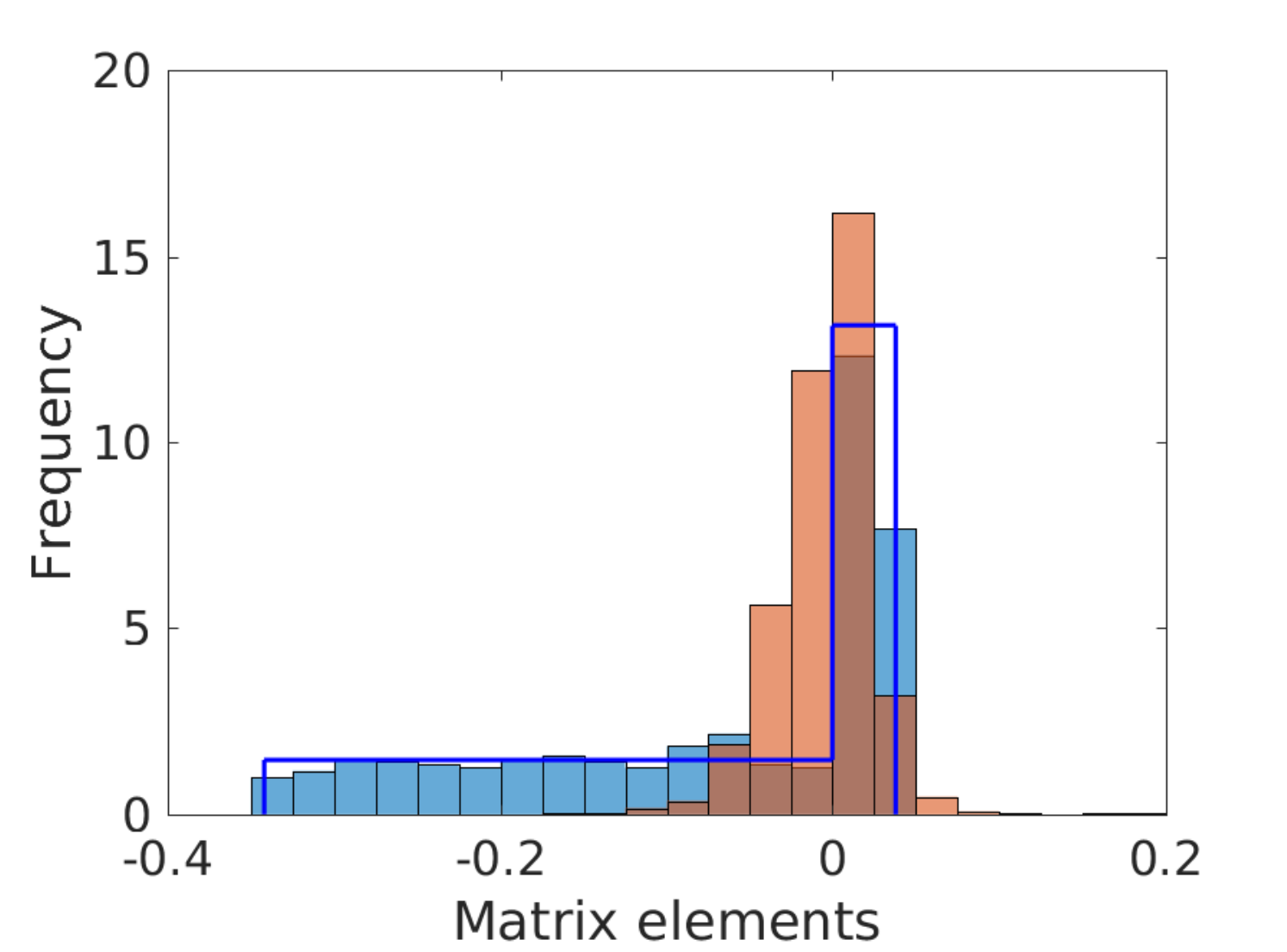}
		}
		\caption{The distribution of the predator-prey interactions is displayed:  blue bars refers to one realization of the initial system (averaging over many realizations yields the analytic profile represented by the blue line and obtained after equations \eqref{histo_analytic}). Red bars photograph the distribution of couplings obtained once the control has been applied.}
		\label{fig_hist_pred_prey_0}
	\end{figure}

\begin{figure}
		\captionsetup{justification=raggedright,singlelinecheck=false}
		\subfigure{
			\includegraphics[width=8cm]{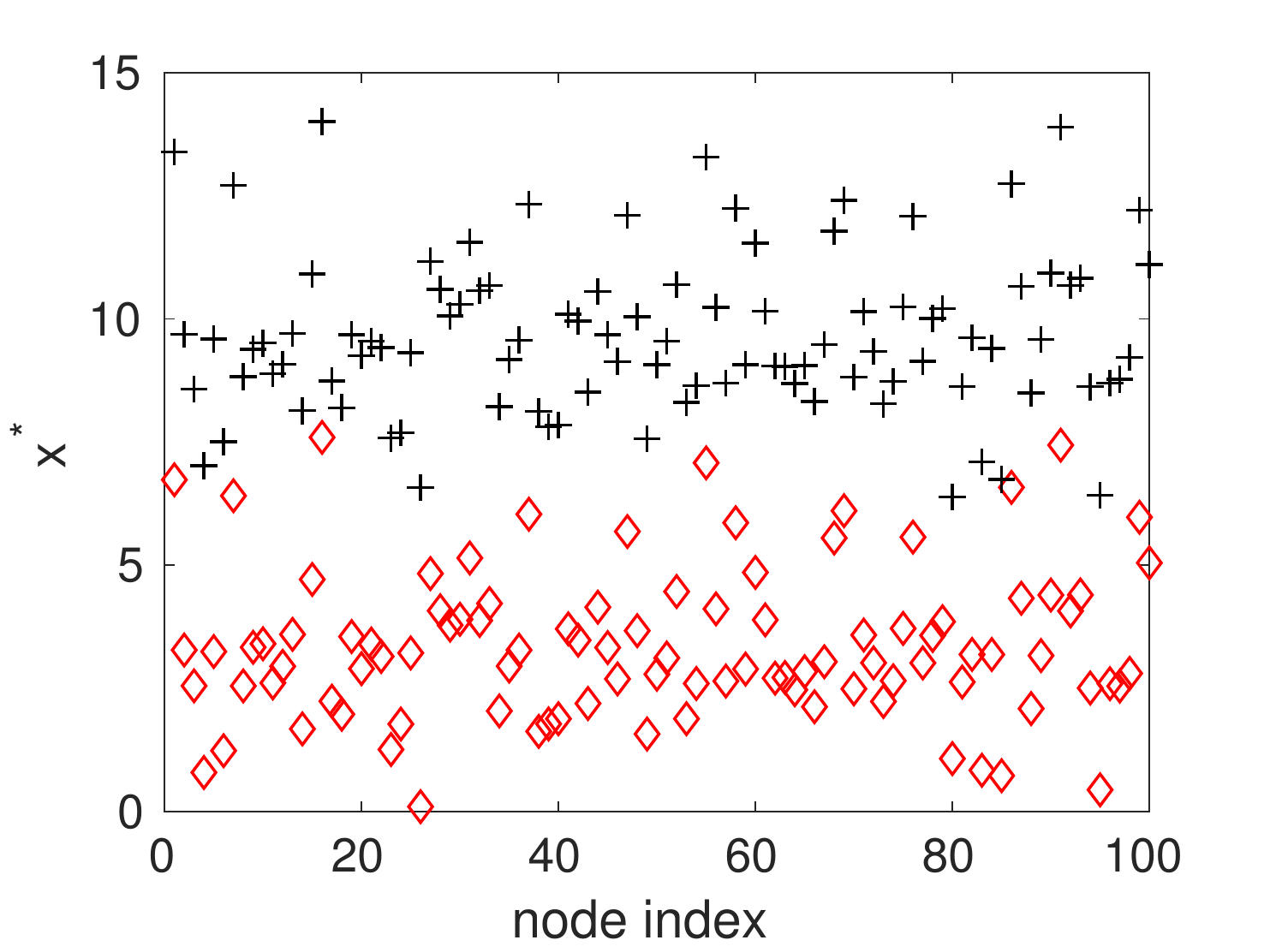}}
		\caption{The equilibrium solution of the controlled system (red diamonds) is different from the trivial (uncoupled) fixed point (black plus symbols) obtained by setting ${x_i^*}'=r_i/s_i$. The index reported in the horizontal axis identifies the species. 		
			\label{fig_fixedpoint}
		}
	\end{figure}

	A different view on the effect of the control method can be gained by looking at Figure \ref{fig_hist_pred_prey_0}. Here, the distribution of the weights associated to predator-prey interactions is plotted before (blue bars and  continuum profile) and after (orange histogram) application of the control. While considering the control procedure which allows the fixed point to change, it is clear that reducing the strength of the couplings proves beneficial for the system stabilization, in agreement with \cite{CoyteSchluterFoster15}. The same observation holds for the other classes of interactions (data not shown). As a final check, we show in Figure \ref{fig_fixedpoint} the new fixed points as obtained with the stabilized matrix $\boldsymbol{A}'$, compared to the trivial fixed point obtained when the matrix of couplings is switched off, i.e. assuming that each species is bound to evolve on a isolated niche.  The method here devised yields a genuinely complex stable equilibrium, which appears to be shaped by the couplings established among interacting units, despite the global tendency to reduce their associated strengths, as revealed in Figure \ref{fig_hist_pred_prey_0}.

	\section{Conclusion}
	\label{sec_concl}
	
	In ecological systems, species interact in pairs and according to well-defined modalities, as e.g. predator-prey, mutualistic or competitive. 
	The dynamics of an ecosystem can be ideally modeled by resorting to a collection of ordinary differential equations for the evolution of the associated mean-field concentrations. This amounts to operating under the deterministic approximation, by deliberately neglecting the role played by finite size fluctuations and disregarding spatial variability. Each
	governing differential equation combines two distinct contributions: (i) local terms, also called reactions, forged to mimic the self-dynamics of the examined species and (ii) non local terms, which stem from the couplings among species. The matrix of interactions is a table where all existing pairwise interactions are stored and categorized. Given a specific ecosystem, the reaction parameters of the model and the associated network of inter-species couplings, it is interesting to speculate on the conditions that allow for the existence of a stable equilibrium. Stability relates to resilience, the ability of the system to withstand changes that would alter its dynamical equilibrium. Working in this context, we here contributed with a novel approach, mathematically grounded on first principles,  to help identifying the topological features that should be possessed by a generic ecological network so as to ensure stability,  hence resilience. To this end we consider a system made up of $N$ interacting species and assume the reaction dynamics to be of the logistic type. We then generate a network of interactions, encoded in a weighted adjacency matrix, which is prone to instability. By rewiring the assigned links, and their associated weights, we can drive the system stable, via two alternative strategies which preserve, or not, the initial fixed point. The method consists of a generalized version of the topological control scheme designed in \cite{Cencetti_etal17}, where diffusive (hence, linear) couplings were instead assumed to drive the exchanges between species. The technique here developed implements minimal modifications to the spectrum of the Jacobian matrix responsible for the stability of the underlying equilibrium and trace these changes back to species-species interactions. The analysis carried out within this operating framework, suggests that  predator-prey interactions exert a stabilizing effects, in qualitative agreement with the conclusion reached in \cite{AllesinaTang12}. Furthermore,  it has been found that a preponderance of weak interactions is beneficial to stability \cite{CoyteSchluterFoster15}.

	\appendix
	
	\section{Couplings in the original system}
	\label{app_orig}

\begin{figure}
		\captionsetup{justification=raggedright,singlelinecheck=false}
		\includegraphics[width=7cm]{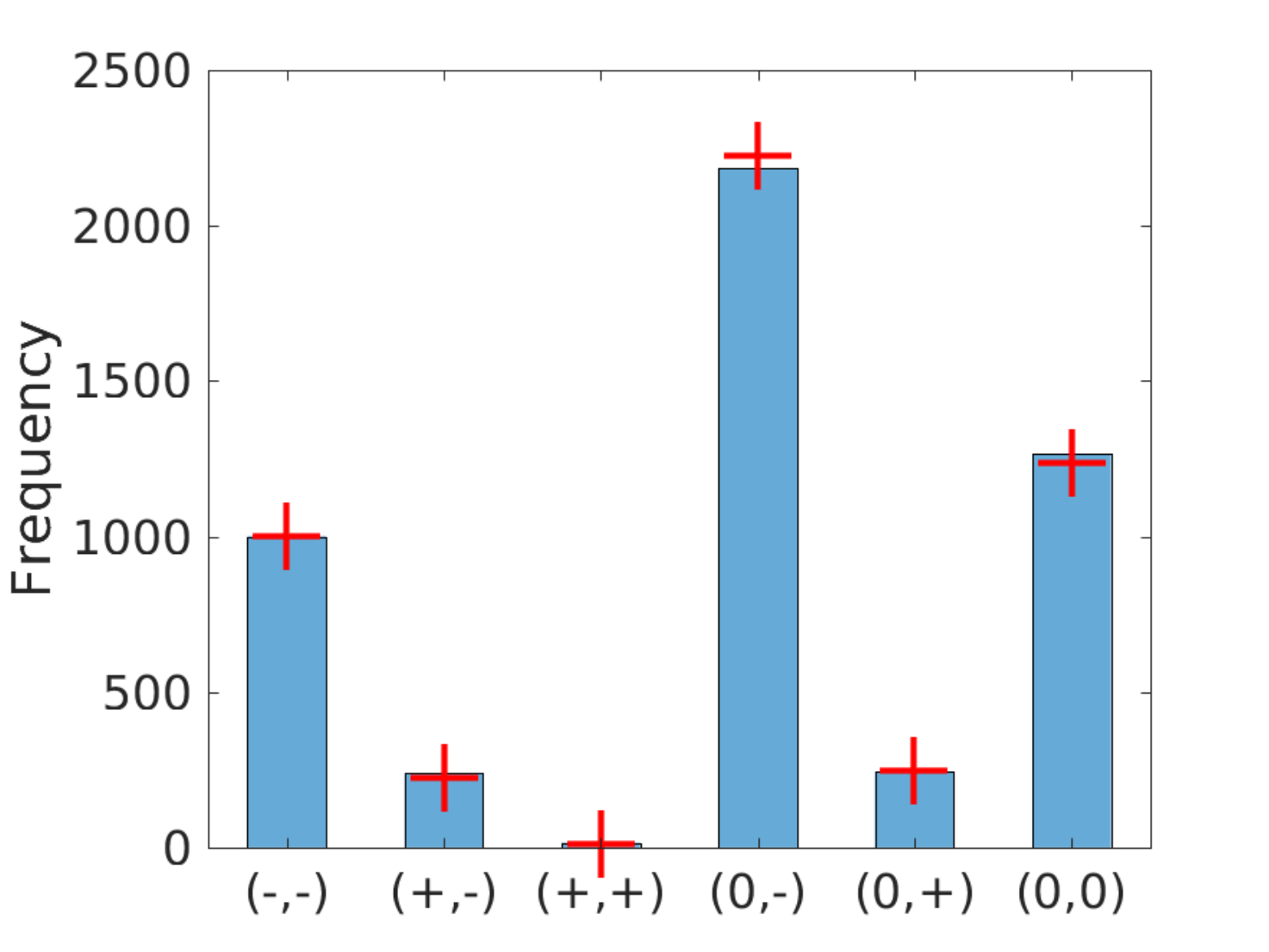}
		\caption{The amount of pairs of the different kinds are shown. Red crosses are analytically computed after equations \eqref{histo_analytic}.  Blue bars are numerically obtained following one individual realization of the scheme discussed above for generating $\boldsymbol A$. Here, $N=100$, $a=0.38$, $p_0=0.5$, $c_0=0.9$. The agreement between the analytic calculation and the numerical implementation is already satisfying (for just one realization) for $N=100$. This observation justifies assuming the analytic abundances as a reference stand for the controlled matrices of interactions (of the same dimension $N=100$) to be compared with.}
		\label{histo_before}
	\end{figure}

We shall here briefly discuss the algorithm that we have employed to generate the initial matrix of interactions $\boldsymbol A$. For each pair $i,j$ (with $i$ and $j$ running from $1$ to $N$) we draw a random number $p$ from a uniform distribution in the interval $[0,1]$. If $p>p_0$ a link that goes from $j$ to $i$ is established. Here $0<p_0<1$ is a free parameter.  
The strength of the link, namely the element $A_{ij}$ of matrix $\boldsymbol A$,  is assigned as follows: we extract a random number from a uniform distribution defined in the compact interval $[-c_0, 1-c_0]$, with $c_0>0$, and multiply the selected number by a scalar amplitude factor $a$. It is immediate to prove, that increasing $c_0$ makes the system progressively more unstable. This follows a direct application of the Gershgorin theorem, mentioned in the main body of the paper. The relative abundance of the pairs, typified as in the main text, can be on average computed and shown to be related to the choices of $p_0$ and $c_0$. The percentage of null entries (no links) of $\bf A$ will be $1-p_0$, the percentage of negative entries $p_0c_0$, while the percentage of positive entries is $p_0(1-c_0)$. Building on this observation, we obtain the following estimate for $f_{(\cdot,\cdot)}$, the frequencies of occurrence of the different classes:	
	\begin{eqnarray}
	f_{(0,0)} &=& (1-p_0)^2\\ \nonumber
	f_{(0,-)}&=& 2p_0(1-p_0)c_0\\ \nonumber
	f_{(0,+)}&=& 2p_0(1-p_0)(1-c_0)\\  \nonumber
	f_{(+,-)}&=& 2p_0^2c_0(1-c_0)\\ \nonumber
	f_{(+,+)}&=& p_0^2(1-c_0)^2\\ \nonumber
	f_{(-,-)}&=& p_0^2c_0^2.\\ \nonumber
	\label{histo_analytic}
	\end{eqnarray}
Each quantity is to be multiplied for a factor $(N^2-N)/2$ to obtain the abundances of elements, which are numerically tested in Figure  \ref{histo_before}. The red crosses depicted in the Figures enclosed in the main body of the paper refer to equations \eqref{histo_analytic}.

	\section{Extending the analysis to account for a generalized nonlinear reaction term.}
	
	The control scheme developed in this paper can also be applied to a more general reaction model. Assume that the logistic dynamics is replaced by a generic nonlinear function $f(x_i,\boldsymbol{r}_i)$:
	
	\begin{equation}
	\dot{x_i}=f(x_i,\boldsymbol{r}_i) + x_i\sum_{j\neq i}A_{ij}x_j
	\end{equation}
	where $\boldsymbol{r}_i$ identifies an arbitrary set of constant parameters.\\
	The associated fixed point is obtained by solving: 
	\begin{equation}
	f(x_i,\boldsymbol{r}_i) + x_i\sum_{j\neq i}A_{ij}x_j=0.
	\end{equation}
	By performing the change of variables $y_i\equiv x_i/x_i^*$ and $B_{ij}=A_{ij}x_j^*$ we obtain:
	\begin{equation}
	\dot{y_i}=\frac{1}{x_i^*}f(y_ix_i^*,\boldsymbol{r}_i) + y_i\sum_{j\neq i}B_{ij}y_j
	\end{equation}
	
	In the new variables, the fixed point equation takes the form:
	\begin{equation}
	\frac{1}{x_i^*}f(x_i^*,\boldsymbol{r}_i) + \sum_{j\neq i}B_{ij}=0
	\end{equation} 
	where use has been made of the condition $y_i^*=1$.\\
	The stability analysis requires introducing a modest perturbation $v_i$ around the fixed point $y_i^*=1$, which amounts to writing $y_i=1+v_i$. The inhomogeneous perturbation $v_i$ will evolve according to: 
	\begin{equation}
	\dot v_i = v_i\bigg[\frac{1}{x_i^*}\frac{\partial f}{\partial y_i}(y_ix_i^*,\boldsymbol{r}_i)\bigg|_{y_i=1} + \sum_{j\neq i}B_{ij}\bigg] + \sum_{j\neq i}B_{ij}v_j\equiv \sum_j C_{ij}v_j.
	\end{equation}
	where the matrix $\boldsymbol C$ is defined as
	\begin{equation}
	C_{ij}\equiv \left\{\begin {array}{ll} 
	B_{ij}\ \ \ \ \ \ \ \ \ \ \ \ \ \ \ \ \ \ \ \ \ \ \text{if}\ i\neq j\\
	\sum_{k\neq i}B_{ik} + \frac{1}{x_i^*}\frac{\partial f}{\partial y_i}(y_ix_i^*,\boldsymbol{r}_i)\bigg|_{y_i=1}\ \ \text{if}\ i=j,\\
	\end{array}\right.
	\end{equation}
	and the stability of the fixed point is ultimately controlled by the sign of the real part of the eigenvalues of $\boldsymbol C$.
	In analogy with the procedure described in the main text, we then define $\boldsymbol D$ as
	\begin{equation}
	D_{ij}\equiv \left\{\begin {array}{ll} 
	B_{ij}\ \ \ \ \ \ \ \text{if}\ i\neq j\\
	\frac{1}{x_i^*}f(x_i^*,\boldsymbol{r}_i)\ \ \text{if}\ i=j,\\
	\end{array}\right.
	\end{equation}
	which is a zero-row-sum matrix, because of the fixed point condition, $\sum_jD_{ij}=\sum_{j\neq i} B_{ij}+\frac{1}{x_i^*}f(x_i^*,\boldsymbol{r}_i)=0$.\\
	We can then modify the matrix $\boldsymbol D$, which hence transforms into $\boldsymbol{D'}$, so as to enforce the desired stability, following the recipe outlined in the main body of the paper. The implemented changes can be interpreted as follows:
	\begin{itemize}
	\item for the off-diagonal entries ($i\neq j$) we impose $D'_{ij} = B'_{ij} = A'_{ij}{x_j^*}'$ which enables to calculate the elements of the controlled interaction matrix. As usual, we can decide to freeze the fixed point to its original value or modify it consistently, while assuming constant the parameters of the model.
		
  \item for the diagonal entries ($i=j$) we impose $D'_{ii}=\frac{1}{{x_i^*}'}f({x_i^*}',\boldsymbol{r}_i')$ where $\boldsymbol{r}_i'$ is the new vector of parameters of the stabilized system.  This latter can be readily obtained by inverting the above equation ($f$ need therefore to  be invertible, with respect to $\boldsymbol r$). Notice that, in general, only a subset of the elements of $\boldsymbol{r}_i$, need to be adjusted. As remarked above, it is alternatively possible to leave the parameters $\boldsymbol{r}_i$  unchanged, and modify $\boldsymbol{x^*}$\\ 
	\end{itemize}

\section{The controlled matrix $\boldsymbol{D'}$ is real and zero-row-sum as $\boldsymbol D$ is}
	\label{app_real_stoch}
	Recall the definition of $\boldsymbol{D'}$:
	\begin{equation}
	\boldsymbol{D}'=\boldsymbol{D}+\boldsymbol \Phi( \boldsymbol{\delta\Lambda})\boldsymbol \Phi^{-1}\equiv\boldsymbol{D}+\boldsymbol{\delta D}
	\end{equation}
	where $\boldsymbol \Phi$ is the matrix whose columns are the eigenvectors $(\boldsymbol{\phi^{(1)}},...,\boldsymbol{\phi^{(N)}})$ of $\boldsymbol D$. Observe that since $\boldsymbol{D}$ is real and zero-row-sum, it is sufficient for our purposes to prove that $\boldsymbol{\delta D}$ exhibits the same properties.\\
	
	{\bf The elements of $\boldsymbol{D'}$ are real.}  Consider the generic entries of $\boldsymbol{\delta D}$:
	\begin{equation}
	(\delta D)_{il}=\sum_j\Phi_{ij}\delta\Lambda^{(j)}(\Phi^{-1})_{jl}.
	\label{d'}
	\end{equation}
	Isolate the real and imaginary parts of every element in Eq.~(\ref{d'}). It is immediate to see that, being $\delta\Lambda^{(j)}$ real $\forall j$, the imaginary part of $(\delta D)_{il}$ reads:
	\begin{equation}
	(\delta D_{Im})_{il} = \sum_j \delta\Lambda_{Re}^{(j)}[(\Phi_{Re})_{ij}(\Phi^{-1}_{Im})_{jl}+(\Phi_{Im})_{ij}(\Phi^{-1}_{Re})_{jl}].
	\label{ImDelta}
	\end{equation}
	
	To match condition $(\delta D_{Im})_{il}=0$, the term on the right hand side of Eq.~(\ref{ImDelta}) should be zero. To prove this, let us first recall that the eigenvalues of a real matrix are either real or complex and come in conjugate pairs. Consider a complex eigenvalue $\Lambda^{(\alpha)}$, by definition  $\boldsymbol{D}\boldsymbol{\phi}^{(\alpha)}= \Lambda^{(\alpha)}\boldsymbol\phi^{(\alpha)}$. Taking the complex conjugate yields 
	$\boldsymbol{D} (\boldsymbol{\phi}^{(\alpha)})^*=(\Lambda^{(\alpha)})^* (\boldsymbol\phi^{(\alpha)})^*$ where $(\cdot)^*$ stands for the complex conjugate and where we have used the condition  $\boldsymbol{D}=  \boldsymbol{D}^*$. Let  the indexes $\alpha$ and $\beta$ to be defined by the  relation $(\Lambda^{(\alpha)})^*=\Lambda^{(\beta)}$. We can immediately conclude that $\boldsymbol\phi^{(\beta)}=(\boldsymbol\phi^{(\alpha)})^*$. Hence
	
	\begin{equation}
	\begin{array}{ll}
	(\Phi_{Re})_{i\alpha}=(\Phi_{Re})_{i\beta}\\
	(\Phi_{Im})_{i\alpha}=-(\Phi_{Im})_{i\beta}
	\end{array}
	\label{ReImPhi}
	\end{equation}
	for each node index $i$ and every pair $(\alpha,\beta)$ of complex conjugate eigenvalues. Consider now the equation
	\begin{equation}
	(\boldsymbol\phi^{-1})^{(\alpha)}\boldsymbol{D}=\Lambda^{(\alpha)}(\boldsymbol\phi^{-1})^{(\alpha)}
	\end{equation}
	with $(\boldsymbol\phi^{-1})^{(\alpha)}$ left eigenvector of $\boldsymbol{D}$, corresponding to the $\alpha$-th row of $\boldsymbol\Phi^{-1}$. Proceeding in analogy with the above, one eventually gets:
	\begin{equation}
	\begin{array}{ll}
	(\Phi^{-1}_{Re})_{\alpha l}=(\Phi^{-1}_{Re})_{\beta l}\\
	(\Phi^{-1}_{Im})_{\alpha l}=-(\Phi^{-1}_{Im})_{\beta l}.
	\end{array}
	\label{ReImPhi_1}
	\end{equation}
	Return now to Eq.~(\ref{ImDelta}). By performing the summation on $j=\alpha$ and $j=\beta$, using Eq.~(\ref{ReImPhi}) and Eq.~(\ref{ReImPhi_1}) and the fact that the corrections $\delta\Lambda^{(\alpha)}$ and $\delta\Lambda^{(\beta)}$ are complex conjugated like the original eigenvalues  $\Lambda^{(\alpha)}$ and $\Lambda^{(\beta)}$, we finally conclude that the terms of the sums in Eq.~(\ref{ImDelta}) cancel in pairs relative to complex conjugate eigenvalues.\\
	The remaining terms in the summation correspond to real eigenvalues. Without loss of  generality, we can always consider the relative left and right eigenvectors to be real, up to a constant scaling factor. In detail, if $\Lambda^{(\gamma)} \in \mathbb{R}$, $(\Phi_{Im})_{i\gamma}$ and $(\Phi^{-1}_{Im})_{\gamma l}$ can always be set equal to zero $\forall i,l$. This implies that all the remaining terms in equation \eqref{ImDelta}  disappear, one by one. Then, summing up, $(\delta D_{Im})_{il}=0\ \forall i,l$.\\

	{\bf $\boldsymbol{D'}$ is a zero-row-sum matrix.}
	 Matrix $\boldsymbol{D}$ zero-row-sum. Hence, the vector $\boldsymbol1$ (with all entries equal to one) is the right eigenvector of $\boldsymbol{D}$ corresponding to $\Lambda^{(1)}=0$:
	\begin{equation}
	\sum_jD_{ij}=0\ \ \ \iff \ \ \ \boldsymbol{D} \boldsymbol1=0.
	\end{equation}
	Recall that the proposed approach, implies a modification of the eigenvalues of matrix $\boldsymbol{D}$, while keeping the eigenvectors unchanged. As a consequence,  vector $\boldsymbol1$ is still solution of the eigenvalue problem. Moreover, the zero eigenvalue is not responsible for the instability and it is therefore preserved upon application of the control scheme. Hence:
	\begin{equation}
	\boldsymbol{D'}\boldsymbol1=0\ \ \iff \ \ \sum_j(D')_{ij}=0
	\end{equation}
	which proves the claim.

		\section{On the conditions of controllability.}
		\label{app_controllability}
		
As explained in the main text, the proposed control method is based on shifting the eigenvalues of a zero-row-sum matrix in the complex plane by applying to their values a real and negative correction (which is identically equal to zero, for the subset of eigenvalues which should be preserved). The goal of this Appendix is to single out the conditions which allow for the control procedure to be effectively implemented.\\
		
Denote by $\mathcal M$ the set of indices corresponding to the eigenvalues which are responsible for the instability and which should be modified by the controller.  Denote by  $\delta\Lambda_i$,  $i\in\mathcal M$, the (real) shift imposed to the selected ensemble of eigenvalues for stabilization. In the sequel, we will assume that the translation 
takes the interested eigenvalues to a constant  value, $R$, smaller than $r_{min}$ (other strategies can clearly be adopted,  consequently altering the analysis reported below):

		\begin{equation}
		(\Lambda_{Re})_i+\delta\Lambda_i=R<r_{min}
		\label{R_app}
		\end{equation}
		In choosing the constant $R$ we must ensure that the applicability constraint \eqref{applicability} is satisfied:
		\begin{equation}
		\tilde s_i-\sum_j\Phi_{ij}\delta\Lambda_j\Phi_{ji}^{-1}\geq0 \ \forall i
		\end{equation}
		Imposing condition \eqref{R},  one gets:
		\begin{equation}
		\tilde s_i - R\sum_{j\in \mathcal M}\Phi_{ij}\Phi_{ji}^{-1} + \sum_{j\in \mathcal M}\Phi_{ij}(\Lambda_{Re})_j\Phi_{ji}^{-1} \equiv k_i - R\sum_{j\in \mathcal M}\Phi_{ij}\Phi_{ji}^{-1}\geq0\ \forall i.
		\label{applicability2}
		\end{equation} 
		where we defined $k_i=\tilde s_i+ \sum_{j\in \mathcal M}\Phi_{ij}(\Lambda_{Re})_j\Phi_{ji}^{-1}$.\\

		If  $k_i - r_{min}\sum_{j\in \mathcal M}\Phi_{ij}\Phi_{ji}^{-1}$ is positive, for each index $i \in [1,N]$, then it is sufficient to set $R=r_{min}$ for achieving stabilization. The complementary situation, in which the above quantity turns out to be negative for some $i$, is more intricate, as we shall clarify hereafter. 
				
		First, it is convenient to sort the node indices in such a way that for indices $i \in [1,n]$, with $n<N$, the quantity $\sum_{j\in \mathcal M}\Phi_{ij}\Phi_{ji}^{-1}$ is positive, while it takes negative values for the remaining indices of the collection, $i \in [n+1,N]$. 
		Recalling that $R$ is bound to be smaller than $r_{min}$, the following inequalities hold:
		\begin{equation}
		k_i - R\sum_{j\in \mathcal M}\Phi_{ij}\Phi_{ji}^{-1} \leq k_i - r_{min}\sum_{j\in \mathcal M}\Phi_{ij}\Phi_{ji}^{-1} \hspace{5mm}\forall i \in [n+1,N]
		\label{app_dis}
		\end{equation} 
		\begin{equation}
		k_i - R\sum_{j\in \mathcal M}\Phi_{ij}\Phi_{ji}^{-1} \geq k_i - r_{min}\sum_{j\in \mathcal M}\Phi_{ij}\Phi_{ji}^{-1} \hspace{5mm}\forall i \in [1,n]
		\label{app_dis2}
		\end{equation} 
		where $R$ is chosen so as to make the term on the left hand side positive. The inequality  \eqref{app_dis} is obviously violated if the expression on the right hand side is negative for at least one $i$ in the interval $[n+1,N]$. In this case, the system cannot be controlled, using the recipe here discussed (which amounts, among the other specificities, to select a constant $R$).  For what concerns the other inequality \eqref{app_dis2}, the righthand term is instead allowed to be negative. Suppose that this happens for indices $i \in [1,\tilde n]$ with $\tilde n \leq n$ (the indices are imagined to be properly sorted). Then, the constant $R$ must be smaller than $r_{min}$, let us say $R = r_{min} - \epsilon$, with $\epsilon>0$. Substituting it into \eqref{applicability2} we obtain a lower bound for $\epsilon$:
		\begin{equation}
		\epsilon \geq \max_{i\in[1,\tilde n]} \biggl(r_{min} - \frac{k_i}{\sum_{j\in \mathcal M}\Phi_{ij}\Phi_{ji}^{-1}}\biggr).
		\end{equation}
		
		Recall again that  $R$ is (our choice) constant.  Assuming $R = r_{min} - \epsilon$, one could eventually loose the controllability condition for indices $i \in [n+1,N]$. The following additional condition needs therefore to be considered:
		\begin{equation}
		k_i - r_{min}\sum_{j\in \mathcal M}\Phi_{ij}\Phi_{ji}^{-1} + \epsilon_{min}\sum_{j\in \mathcal M}\Phi_{ij}\Phi_{ji}^{-1} \geq 0
		\end{equation}
		thus resulting in  an upper bound for $\epsilon$: 
		\begin{equation}
		\epsilon \leq \min_{i\in[n+1,N]} \biggl(r_{min} - \frac{k_i}{\sum_{j\in \mathcal M}\Phi_{ij}\Phi_{ji}^{-1}}\biggr).
		\end{equation}
		Another necessary condition for controllability is then found by imposing that the upper bound for $\epsilon$ is larger than the lower bound (both positive):
		\begin{equation}
		0<\max_{i\in[1,\tilde n]} \biggl(r_{min} - \frac{k_i}{\sum_{j\in \mathcal M}\Phi_{ij}\Phi_{ji}^{-1}}\biggr) 
		\leq
		\min_{i\in[n+1,N]} \biggl(r_{min} - \frac{k_i}{\sum_{j\in \mathcal M}\Phi_{ij}\Phi_{ji}^{-1}}\biggr)
		\end{equation}
		
		which coincides with the equations reported in the main body of the paper.

	\section{Controlling without modifying the fixed point}
	
\begin{figure}
		\captionsetup{justification=raggedright,singlelinecheck=false}
		\subfigure[]{
			\includegraphics[width=8cm]{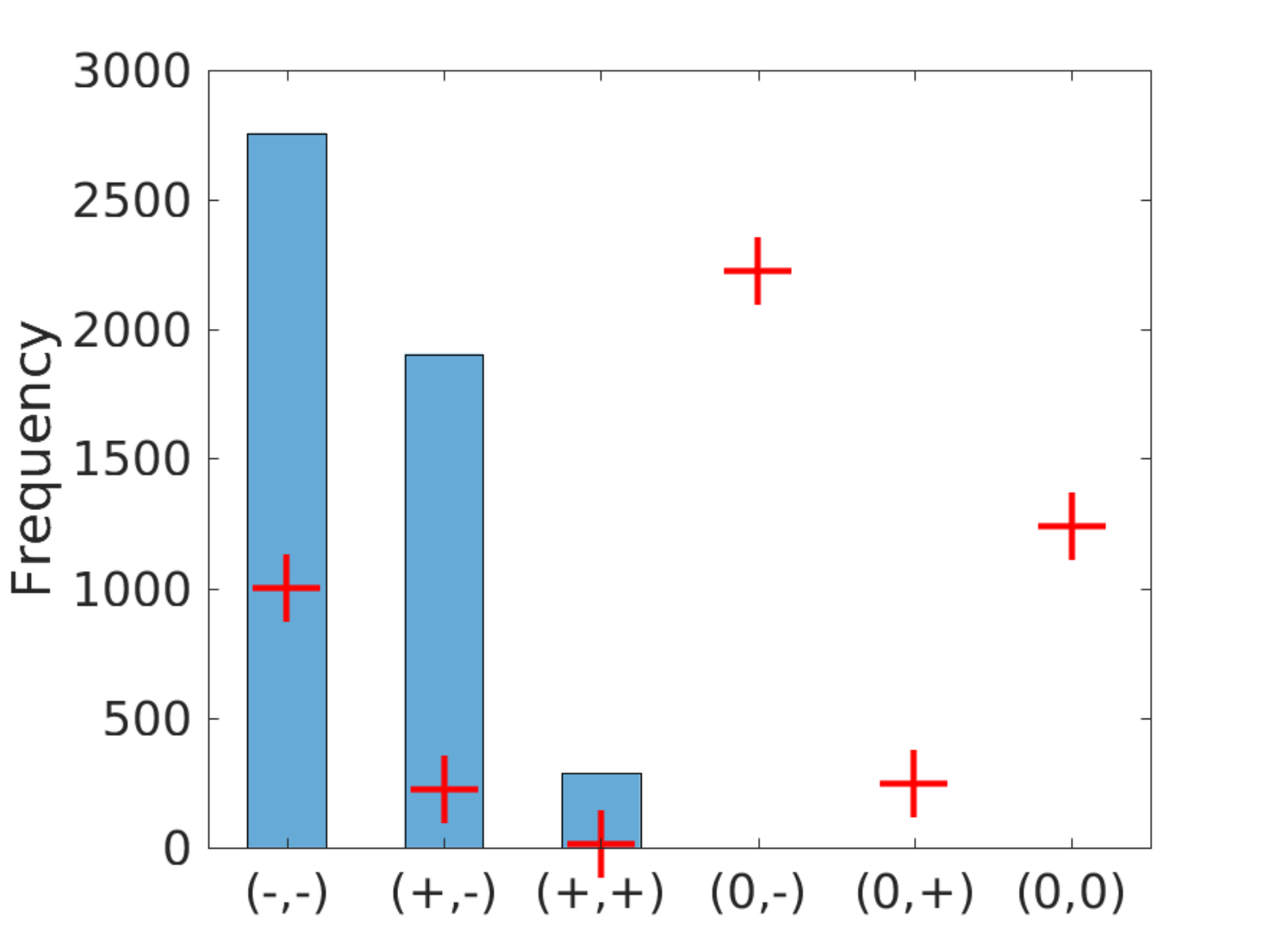}
			}
		\subfigure[]{
			\includegraphics[width=8cm]{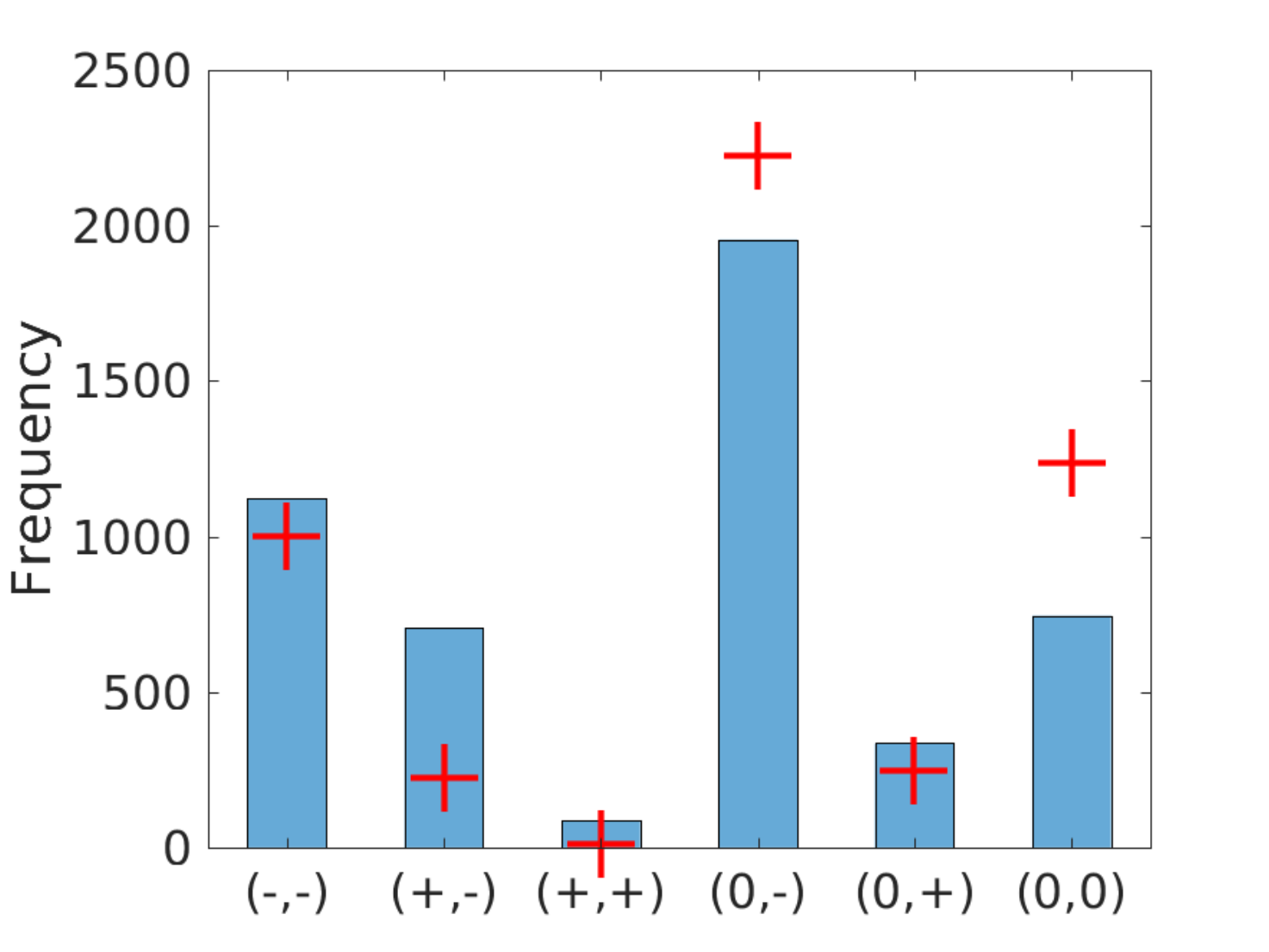}
			}
		
		\caption{Abundances of different types of couplings between species, as explained in caption of Figure \ref{ecology_histo}. Here the blue bars refer to the system controlled upon application of the strategy which leaves the fixed point to its original value, while the parameters $\boldsymbol s$ are modified. The Figure in panel (b) reports the coupling abundances after a cut-off of 0.05 has been applied.}
		\label{1_ecology_histo}
	\end{figure}

\begin{figure}
		\captionsetup{justification=raggedright,singlelinecheck=false}
		\subfigure[]
		{\includegraphics[width=10cm]{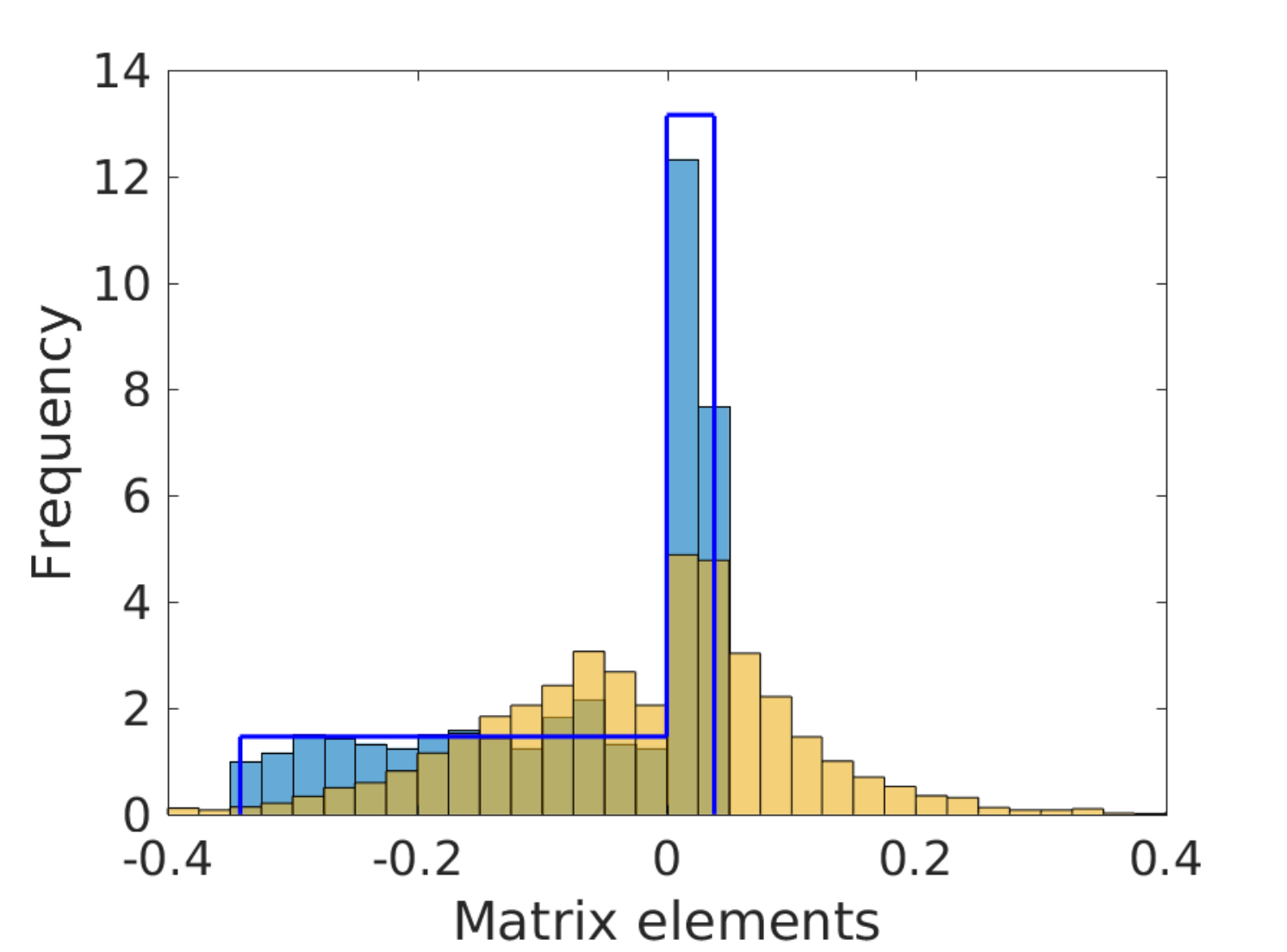}
		}
		\caption{Distribution of predator-prey interactions: the blue bars and the blue line refer to one realization of the initial system, as explained in the caption of Figure \ref{fig_hist_pred_prey_0}. Yellow bars represent the different couplings in matrix $\boldsymbol{A'}$ as obtained from controlling the system leaving the fixed point unchanged. In this case, strong predator-prey links are generated and the distribution of positive vs. negative weights symmetrized}
		\label{fig_hist_pred_prey_1}
	\end{figure}

\begin{figure}
			\captionsetup{justification=raggedright,singlelinecheck=false}
			\subfigure{
				\includegraphics[width=8cm]{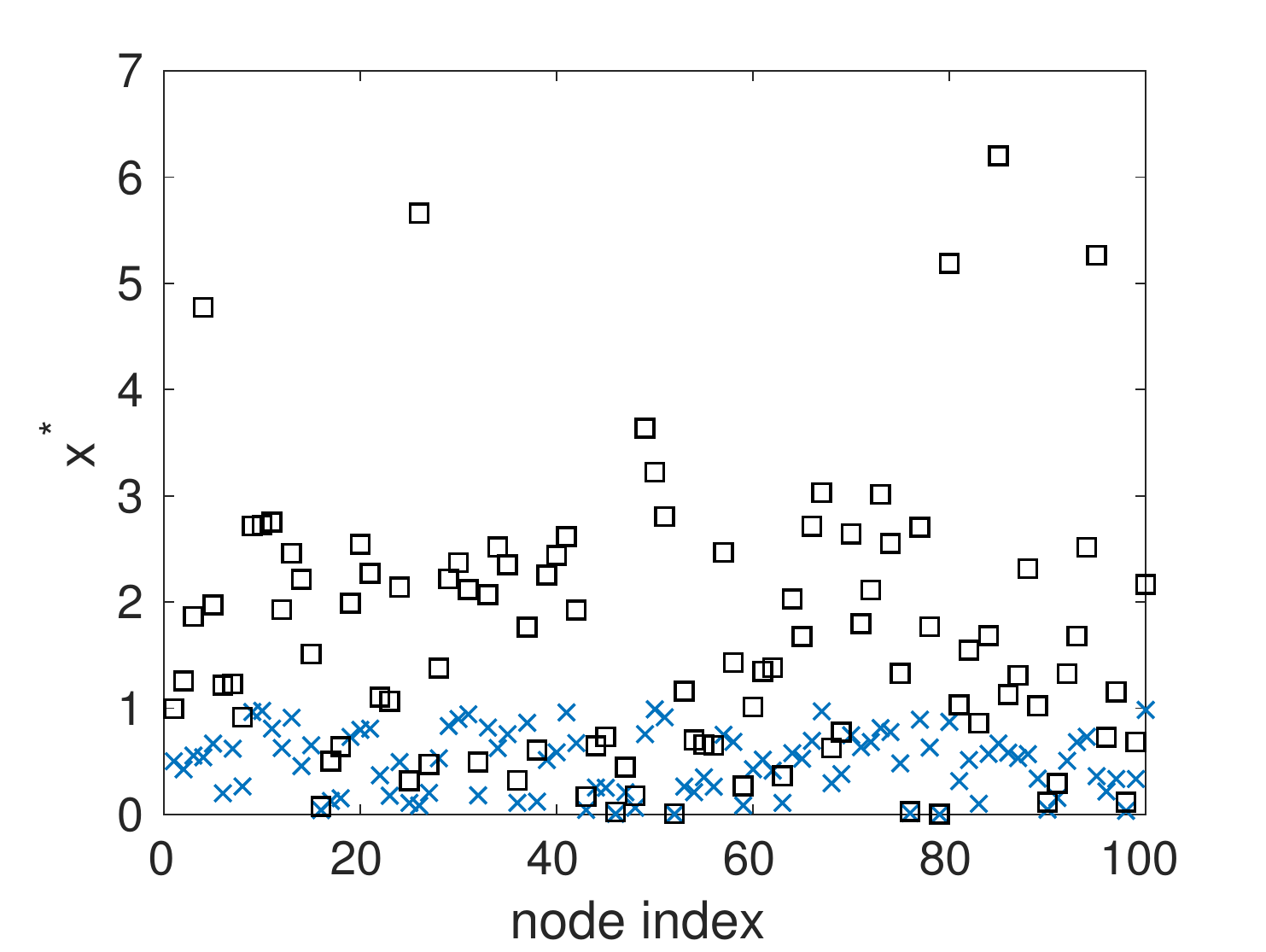}}
			\caption{In analogy with Figure \ref{fig_fixedpoint}, the fixed point of the controlled system is reported with (blue) crosses and compared with the  fixed point, (black) squares, that one would obtain for the uncoupled dynamics (${x_i^*}=r_i/s_i'$).  		
				\label{fig_triv_1}
			}
		\end{figure}

	 The Figures appearing in this Appendix illustrate the results obtained for the control strategy where the fixed point is kept unchanged while the inverse carrying capacity vector $\boldsymbol s$ is modified together with the weights of the interaction matrix. Figures \ref{fig_hist_pred_prey_1},  \ref{fig_triv_1} and \ref{1_ecology_histo} represent the analogue of Figures, respectively  \ref{ecology_histo}, \ref{fig_hist_pred_prey_0} and \ref{fig_fixedpoint}, already discussed in the main body of the paper, which are obtained with the alternative control strategy.

	\

\addcontentsline{toc}{chapter}{Bibliography}
\bibliographystyle{ieeetr} 
\bibliography{MyBib}

\end{document}